\documentclass[10pt, aps, prd, showpacs, twocolumn, notitlepage, nofootinbib, preprintnumbers]{revtex4-1}
\usepackage{amsmath,amssymb}
\usepackage{graphicx}
\usepackage[dvipsnames]{xcolor}  
\usepackage{hyperref} 
\usepackage{lipsum}

\usepackage{tikz}
\usetikzlibrary{decorations.pathmorphing,decorations.markings, shapes.geometric,shapes.misc,calc}
\tikzstyle{gluon}=[decorate, decoration={coil,aspect=0.8, amplitude=1.5pt,  segment length=3pt}]

\newcommand{\dd}{\ensuremath{\text{d}^2}}

\newcommand{\wvec}{{\underline{w}}}
\newcommand{\xvec}{{\underline{x}}}
\newcommand{\Xvec}{{\underline{X}}}
\newcommand{\yvec}{{\underline{y}}}
\newcommand{\Yvec}{{\underline{Y}}}

\newcommand{\zv}{{\underline{0}}}
\newcommand{\ov}{{\underline{1}}}
\newcommand{\tv}{{\underline{2}}}

\newcommand{\bpsi}{\ensuremath{{\bar{\psi}}}}
\newcommand{\thalf}{\ensuremath{{\tfrac{1}{2}}}}

\newcommand{\cald}{\ensuremath{{\mathcal{D}}}}
\newcommand{\calk}{\ensuremath{{\mathcal{K}}}}

\newcommand{\calw}{\ensuremath{{\mathcal{W}}}}
\newcommand{\hcalo}{\ensuremath{{\hat{\mathcal{O}}}}}

\newcommand{\req}[1]{{Eq.~(\ref{#1})}}

\newcommand{\ColorOnline}{{(Color online.)}}

\def\eq#1{{Eq.~(\ref{#1})}}
\def\fig#1{{Fig.~\ref{#1}}}
\newcommand{\ben}{\begin{eqnarray*}}
\newcommand{\een}{\end{eqnarray*}}
\newcommand{\un}[1]{\underline{#1}}

\newcommand{\pd}{\partial}

\newcommand{\tr}{\mbox{tr}}

\newcommand{\as}{\alpha_s}

\newcommand{\ul}[1]{\underline{#1}}

\newcommand{\Tr}{\mathrm{Tr}}

\begin{document}
\title{Helicity-dependent generalization of the JIMWLK evolution}

\author{Florian Cougoulic} 
         \email[Email: ]{cougoulic.1@osu.edu}
         \affiliation{Department of Physics, The Ohio State
           University, Columbus, OH 43210, USA}
           
\author{Yuri V. Kovchegov} 
         \email[Email: ]{kovchegov.1@osu.edu}
         \affiliation{Department of Physics, The Ohio State
           University, Columbus, OH 43210, USA}

\begin{abstract}
We generalize the JIMWLK evolution equation to include the small-$x$ evolution of helicity distributions. To derive helicity JIMWLK, we use the method devised by A.H. Mueller in \cite{Mueller:2001uk} for the usual unpolarized JIMWLK, and apply it to the helicity evolution equations derived in \cite{Kovchegov:2015pbl,Kovchegov:2016zex,Kovchegov:2017lsr,Kovchegov:2018znm}. We obtain a functional evolution equation for the generalized JIMWLK weight functional. The functional now is a sum of a helicity-independent term and a helicity-dependent term. The kernel of the new evolution equation consists of the standard eikonal leading-order JIMWLK kernel plus new terms containing derivatives with respect to the sub-eikonal quark and gluon fields. The new kernel we derive can be used to generate the standard JIMWLK evolution and to construct small-$x$ evolution equations for flavor-singlet operators made of any number of light-cone Wilson lines along with one Wilson line containing sub-eikonal helicity-dependent local operator insertions (a ``polarized Wilson line").   
\end{abstract}

\maketitle
\tableofcontents

\section{Introduction}

In recent years, small-$x$ asymptotics of transverse momentum dependent parton distribution functions (TMD PDFs or TMDs for short) received considerable attention in the literature \cite{Boer:2006rj,Boer:2002ij,Boer:2008ze,Balitsky:2015qba,Altinoluk:2014oxa,Kovchegov:2013cva,Altinoluk:2015gia,Boer:2015pni,Kovchegov:2015pbl,Hatta:2016aoc,Dumitru:2015gaa,Chirilli:2018kkw,Kovchegov:2019rrz,Boussarie:2019icw}. In a series of papers \cite{Kovchegov:2015pbl,Kovchegov:2016zex,Kovchegov:2016weo,Kovchegov:2017jxc,Kovchegov:2017lsr,Kovchegov:2018znm} the small-$x$ evolution equations for the quark and gluon helicity distributions were constructed by finding sub-eikonal corrections to the eikonal shock wave formalism. The resulting equations are written for correlators of infinite light-cone Wilson lines, along with the so-called `polarized Wilson lines', operators consisting of light-cone Wilson lines with the sub-eikonal helicity-dependent local operator insertions \cite{Kovchegov:2017lsr,Kovchegov:2018znm}. These equations are the helicity analogue of the first equations in the Balitsky hierarchy \cite{Balitsky:1995ub,Balitsky:1998ya} for the unpolarized small-$x$ evolution. Similar to the unpolarized case of the Balitsky--Kovchegov (BK) equation \cite{Balitsky:1995ub,Balitsky:1998ya,Kovchegov:1999yj,Kovchegov:1999ua}, the helicity evolution equations of \cite{Kovchegov:2015pbl} close in the large-$N_c$ limit, where $N_c$ is the number of quark colors. (That is, one obtains the same correlator entering the expressions on both sides of the equation, which can then be solved for that correlator.) Unlike the unpolarized case, the helicity evolution equations also close in the large-$N_c \,\&\, N_f$ limit (where $N_f$ is the number of quark flavors), since in the case of helicity evolution quark loops contribute already at the leading order. No closed integro-differential equations can be obtained for helicity evolution in the shock wave formalism in the case of arbitrary (not large) $N_c$ and $N_f$. In addition, the helicity evolution equations were constructed in \cite{Kovchegov:2015pbl,Kovchegov:2017lsr,Kovchegov:2018znm} for the so-called polarized dipole operators, consisting of a trace of one polarized and one unpolarized (standard) Wilson lines. It would be desirable to generalize this approach to the case of any sub-eikonal helicity-dependent operator made out of an arbitrary number of unpolarized Wilson lines and one polarized Wilson line. 

In the unpolarized case, the generalization of the BK evolution to all $N_c$ is accomplished by employing the Jalilian-Marian--Iancu--McLerran--Weigert--Leonidov--Kovner
(JIMWLK)
\cite{Jalilian-Marian:1997dw,Jalilian-Marian:1997gr,Weigert:2000gi,Iancu:2001ad,Iancu:2000hn,Ferreiro:2001qy}
functional evolution equation. JIMWLK prescription also allows one to write a small-$x$ evolution equation for an arbitrary operator made out of light-cone Wilson lines. In addition, JIMWLK equation can be solved numerically \cite{Weigert:2000gi,Rummukainen:2003ns,Schlichting:2014ipa}, allowing one to construct small-$x$ asymptotics for a variety of Wilson-line correlators. It appears beneficial to extend the JIMWLK-type treatment to helicity evolution at small $x$.

In this paper we construct a helicity analogue of JIMWLK evolution \cite{Jalilian-Marian:1997dw,Jalilian-Marian:1997gr,Weigert:2000gi,Iancu:2001ad,Iancu:2000hn,Ferreiro:2001qy}, or, more precisely, a generalization of the JIMWLK evolution that includes helicity dependence. This equation should be solvable numerically, allowing for a numerical evaluation of the small-$x$ asymptotics of helicity distribution beyond previously available approximations, large $N_c$ and large $N_c \,\&\, N_f$ \cite{Kovchegov:2015pbl,Kovchegov:2016zex,Kovchegov:2016weo,Kovchegov:2017jxc,Kovchegov:2017lsr,Kovchegov:2018znm}.
Helicity JIMWLK equation would additionally allow one to construct the small-$x$ evolution of operators beyond the polarized dipole case, such as the polarized color quadrupole, an operator consisting of a trace of one polarized Wilson line and three standard Wilson lines.

We will use the approach developed by A.H. Mueller for derivation of JIMWLK equation \cite{Mueller:2001uk}, also described in \cite{Kovchegov:2012mbw}. The derivation \cite{Mueller:2001uk} is not the original way of deriving JIMWLK evolution. Rapidity evolution in high energy scattering can be described by two different but equivalent approaches. One either evolves the target weight functional as done in the the original derivation of JIMWLK \cite{Jalilian-Marian:1997dw,Jalilian-Marian:1997gr} (in general a difficult task) or the projectile wave function \cite{Mueller:1993rr,Mueller:1994jq,Mueller:1994gb} (can be done for test operators consisting of Wilson lines \cite{Balitsky:1995ub,Balitsky:1998ya,Kovchegov:1999yj,Kovchegov:1999ua,Mueller:2001uk}). This equivalence allows us to extract the helicity JIMWLK evolution kernel from the rapidity evolution equations we derive here for the test operators containing polarized Wilson lines, explicit expressions for which were constructed in \cite{Kovchegov:2017lsr,Kovchegov:2018znm}. We then convert this helicity JIMWLK kernel into the evolution kernel for the target weight functional.

A new feature of helicity-dependent distributions, compared to the unpolarized ones, is the use of Wilson lines beyond the eikonal approximation \cite{Kovchegov:2018znm,Kovchegov:2017lsr}.
In order to be sensitive to the target helicity, the probe must interact with the target in a sub-eikonal way at least once. In particular, at the leading sub-eikonal approximation (order-$1/s$ with $s$ the center-of-mass energy squared) one focuses on exactly one or two sub-eikonal interaction(s) between the probe and the target: either an exchange of a single sub-eikonal $t$-channel gluon, or two $t$-channel quarks  \cite{Kovchegov:2018znm,Kovchegov:2017lsr}. As a consequence, the probe is not only sensitive to the large eikonal background gluon field generated by the target but also to sub-eikonal gluon and quark background fields, one of them being the chromo-magnetic background field (the $F^{12}$ component of the gluon field strength tensor).
This property will be present in the evolution kernel acting on the test operator representing the probe, and thus will be inherited by the evolution kernel acting on the weight functional. It will allow us to write a helicity dependent evolution equation for the target weight functional depending on the eikonal gluon field along with the sub-eikonal quark and gluon fields, thus generalizing the eikonal JIMWLK evolution to include helicity-dependent effects. This paper focuses only on the flavor singlet helicity evolution, while the flavor non-singlet case is left for a future project.

The paper is structured as follows.
In Sec.~\ref{Sec:2}, we briefly review the method used in \cite{Mueller:2001uk}. We then adapt the method to the helicity-dependent case and derive the main tools used in this paper. The kernel for the flavor singlet helicity evolution of a test operator consisting of two Wilson lines in any irreducible representation (irrep) of the gauge group SU($N_c$) (with one of them being polarized) is constructed in Sec.~\ref{Sec:3a} based on the calculation of helicity-dependent evolution of polarized dipoles carried out in \cite{Kovchegov:2018znm}. This kernel is successfully cross checked with the results of Bartels, Ermolaev and Ryskin \cite{Bartels:1996wc} at the level of the ladder approximation in Sec.~\ref{Sec:3b}. In the Sec.~\ref{Sec:4}, we obtain the helicity-dependent generalization of the JIMWLK equation for the target weight functional. This main result of this work is given in \eq{EQ:Helicity_evolution_equation} with the kernel given by  \req{Eq:Full_helicity_Kernel}.  We also underline some properties expected for the weight functional solving this equation and for the initial condition of this evolution equation. We conclude in Sec.~\ref{Sec:5} by summarizing our main results and by providing an outlook for the all-$N_c$ evolution of the flavor non-singlet helicity TMDs (the large-$N_c$ version of the flavor non-singlet helicity evolution was constructed in \cite{Kovchegov:2016zex}). Throughout the paper we will use the light-cone coordinates defined by $x^\pm = (x^0 \pm x^3)/\sqrt{2}$ while transverse vectors will be defined by ${\un x} = (x^1, x^2)$ with $x_\perp = |{\un x}|$.


\section{The method and useful relations}
\label{Sec:2}

\subsection{Brief overview of the method}
\label{Sec:2A}

Here we review the method of \cite{Mueller:2001uk} used for re-deriving the unpolarized JIMWLK evolution. Let us start with the rapidity-dependent expectation value of an arbitrary operator $\hcalo_\alpha$ given by \cite{Mueller:2001uk,Kovchegov:2012mbw}
\begin{equation}\label{Oexp}
\langle \hcalo_\alpha \rangle_Y = \frac{\int \cald\alpha \ \hcalo_{\alpha} \, {\cal W}_Y[\alpha] }{\int \cald \alpha \ {\cal W}_Y[\alpha]}
\end{equation}
and agree for simplicity that the weight functional ${\cal W}_Y[\alpha]$ is normalized to one,
\begin{equation}\label{Wnorm}
\int \cald \alpha \ {\cal W}_Y[\alpha] = 1 .
\end{equation}
The expectation value \eqref{Oexp} is taken in the target state. 
Here $\alpha (x^-, {\un x}) \equiv A^+ (x^-, {\un x})$, where $A^+$ is the gluon field in the Lorenz gauge $\partial_\mu A^\mu=0$ or in the $A^- =0$ light-cone gauge of the projectile. (Throughout this paper we assume that the target proton or nucleus is moving in the light-cone plus direction, while the projectile is moving in the minus direction.) We have employed the classical equations of motion for the gluon field in the Lorenz gauge,
\begin{equation}
\Box A^\mu_a = \delta^{\mu+} \rho_a,
\end{equation}
in order to replace the usual functional integration over the color charge density $\rho^a(x^-,\xvec)$ by the integration over the field $\alpha^a(x^-,\xvec)$. To construct an evolution equation for the weight functional ${\cal W}_Y[\alpha]$, one first has to derive the evolution equation for the test operator
\begin{equation}
\label{EQ:K_acting_on_O}
\partial_Y \langle \hcalo_\alpha \rangle_Y =
\int \cald \alpha \left( \calk \cdot \hcalo_\alpha \right) {\cal W}_Y[\alpha]
\end{equation}
where $\calk$ is the evolution kernel for the test operator $\hcalo$ generating an infinitesimal step in rapidity $\text{d}Y$. 
On the other hand, the rapidity evolution for the test operator can be obtained by simply differentiating \eq{Oexp} with respect to $Y$  \cite{Mueller:2001uk}, 
\begin{equation}
\label{EQ:Partial_acting_on_W}
\partial_Y \langle \hcalo_\alpha \rangle_Y =
\int \cald \alpha \ \hcalo_\alpha\ \partial_Y {\cal W}_Y[\alpha].
\end{equation}
Equating the two expressions on the right of \req{EQ:K_acting_on_O} and \req{EQ:Partial_acting_on_W}, while making the evolution kernel $\calk$ act on ${\cal W}_Y$ in the former through integration by parts, yields an evolution equation for ${\cal W}_Y$ \cite{Mueller:2001uk},
\begin{equation}\label{Wevol}
\partial_Y {\cal W}_Y[\alpha] = \calk \cdot {\cal W}_Y [\alpha].
\end{equation}
For the unpolarized small-$x$ evolution, \eq{Wevol} is the JIMWLK equation \cite{Jalilian-Marian:1997dw,Jalilian-Marian:1997gr,Weigert:2000gi,Iancu:2001ad,Iancu:2000hn,Ferreiro:2001qy}.

Following \cite{Mueller:2001uk} consider the test operator $\hcalo_\alpha$ being made of two light-cone Wilson lines at positions $\xvec_0$ and $\xvec_1$ in irreducible representations (irrep's) $R_0$ and $R_1$. In order to work at finite $N_c$, we do not take any contraction over the color of the two Wilson lines and write
\begin{equation}
\label{EQ:O_dipole}
\hcalo_{\ov,\zv} = W^{(R_0) \, \dagger}_\zv \otimes W^{(R_1)}_\ov .
\end{equation}
In our notation, light-cone Wilson line in an irrep $R$ is defined by
\begin{align}
W^{(R)}_{\un x} [b^-, a^-] \equiv \mbox{P} \, e^{ i g \int\limits_{a^-}^{b^-} d x^- \, t_R^a \, \alpha^{a} (x^+ =0, x^-, {\un x})}
\end{align}
with the infinite light-cone Wilson line denoted by
\begin{align}
W^{(R)}_{\un x} \equiv W^{(R)}_{\un x} [\infty, - \infty] .
\end{align}
Here $t_R^a$ are SU($N_c$) generators in representation $R$. 

\begin{figure}
\includegraphics[scale=.95]{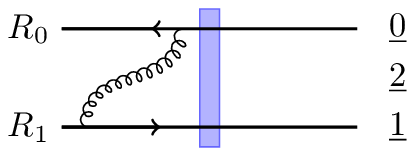}
\includegraphics[scale=.95]{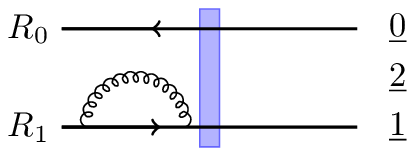}
\includegraphics[scale=.95]{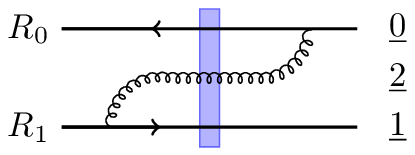}
\includegraphics[scale=.95]{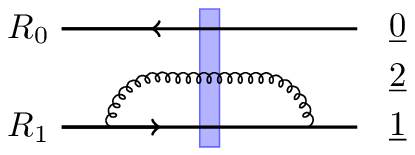}
\includegraphics[scale=.95]{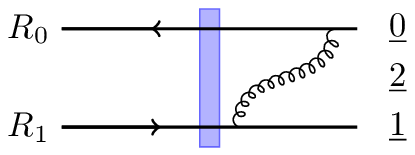}
\includegraphics[scale=.95]{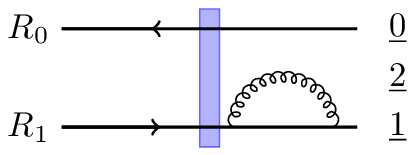}
\caption{Non-exhaustive list of the diagrams contributing to the rapidity evolution of the test operator $\hcalo_{\ov,\zv}$. The blue (gray) rectangle indicates the target shock wave localized at $x^- = 0$. Within the shock wave multiple-rescattering occurs between the test operator $\hcalo_\alpha$ and the target. Remaining diagrams are easily obtained by exchanging $W^{(R_0)\dagger}_\zv \leftrightarrow W^{(R_1)}_\ov$. \ColorOnline 
}
\label{FIG:Evolution_diagrams}
\end{figure}

One step in the evolution of the test operator $\hcalo_{\ov,\zv}$ involves the diagrams depicted in \fig{FIG:Evolution_diagrams}. Consider the first diagram of the second line, generated from the operator $\hcalo_{\ov,\zv}$ after an infinitesimal increase of rapidity $\text{d}Y$,
\begin{equation}
\label{EQ:One_diagr_evolution}
- \frac{\alpha_s}{\pi^2}\int \dd x_2 \, \frac{\xvec_{21}\cdot\xvec_{20}}{x_{21}^2\ x_{20}^2} \ U_\tv^{ba} \left[  W^{(R_0) \, \dagger}_\zv t^b_{R_0} \otimes  W^{(R_1)}_\ov t^a_{R_1} \right]
\end{equation}
with $U_\tv$ denoting the infinite adjoint Wilson line at the transverse position $\xvec_2$. The transverse vectors are defined by ${\un x}_{ij} = {\un x}_i - {\un x}_j$ with $x_{ij} = |{\un x}_{ij}|$. There are two difficulties we need to take care of here: (i) the color generators involved depend on the representation of the Wilson line and (ii) the positions of the two gluon vertex operators with respect to the shock-wave is not the same in every diagram of \fig{FIG:Evolution_diagrams}. The two issues are solved by introducing a functional derivative of a Wilson line with respect to the background field $\alpha$,
\begin{align}
\label{EQ:WilsonLine_Derivative_with respect to_alpha}
&\frac{\delta\ W^{(R)}_\xvec}{\delta\,\alpha^a(y^-,\yvec)} = \,ig \,  \delta^{(2)}(\yvec - \xvec)\   \nonumber \\ 
&\qquad \times W^{(R)}_\xvec [\infty , y^-] \,  t^a_{R}\, W^{(R)}_\xvec [y^- , -\infty]
\end{align}
involving semi-infinite Wilson lines in irrep $R$. Anticipating $x^-$-ordering of small-$x$ evolution we observe that a Wilson line that does not cross the shock-wave (placed at $x^-=0$) is simply an identity in the corresponding color space, while a semi-infinite Wilson line which does cross the shock wave can be formally completed to an infinite Wilson line without changing its value. These considerations allow us to simplify \req{EQ:WilsonLine_Derivative_with respect to_alpha} to
\begin{equation}
\label{EQ:WilsonLine_Derivative_with respect to_alpha_2}
\frac{\delta\ W^{(R)}_\xvec}{\delta\,\alpha^a(y^-,\yvec)} = 
ig \,  \delta^{(2)}(\yvec - \xvec)
\begin{cases}
W^{(R)}_\xvec\, t_R^a & \!\! \text{for } y^- < 0, \\
t_R^a\, W^{(R)}_\xvec & \!\! \text{for } y^- > 0.
\end{cases}
\end{equation}
We readily see that replacing each color generator in the diagrams of \fig{FIG:Evolution_diagrams} by functional derivative solves both issues (i) and (ii). This simple treatment has several advantages. The sum over all vertex positions with respect to the shock wave will be automatically taken care of by the Leibniz rule (product rule) for functional derivatives; as a consequence, the generalization to any number of Wilson lines used to define $\hcalo_\alpha$ is already accomplished.
The color representations of the Wilson lines used to define the test operator $\hcalo_\alpha$ do not matter anymore since \req{EQ:WilsonLine_Derivative_with respect to_alpha_2} holds for any irrep $R$.
This shows how the evolution kernel obtained from the dipole evolution applies to a broad class of test operators consisting of any number of Wilson lines in any irreps.

For completeness, let us write down the unpolarized small-$x$ evolution kernel acting on $\hcalo_\alpha$ in \eq{EQ:K_acting_on_O} we obtained by this method, 
\begin{widetext}
\begin{equation}
\label{EQ:JIMWLK_Kernel}
\mathcal{K}_{JIMWLK} \equiv \frac{\alpha_s}{\pi^2}\int \dd x_\perp \, \dd y_\perp \, \dd w_\perp \ 
\frac{(\xvec-\wvec)\cdot(\yvec-\wvec)}{|\xvec-\wvec|^2\ |\yvec-\wvec|^2} \left( U_\wvec - \frac{U_\xvec + U_\yvec}{2}\right)^{ba}  
\frac{(ig)^{-2}  \ \delta^2}{\delta\,\alpha^a(x^-<0,\xvec)\ \delta\,\alpha^b(y^->0,\yvec)} .
\end{equation}
\end{widetext}
One can show that this kernel is equivalent to the kernels used in \cite{Kovner:2005jc,Kovner:2005aq} or in \cite{Mueller:2001uk,Kovchegov:2012mbw}, while being perhaps a little more compact.
We are now in a position to state that the JIMWLK equation for the evolution of the target weight functional is given by \eq{Wevol} with the kernel from \eq{EQ:JIMWLK_Kernel}.
Notice that after the integration by parts in \eq{EQ:K_acting_on_O} the functional derivatives do not act on the adjoint Wilson lines in the parenthesis of \req{EQ:JIMWLK_Kernel}. 
This simply follows from the fact that the structure constants of SU($N_c$) are fully anti-symmetric. Thus the kernel of \eq{EQ:JIMWLK_Kernel} remains the same after the integration by parts and enters \eq{Wevol} unaltered.


\subsection{Adapting the method to helicity-dependent scattering}
\label{Sec:2B}

\begin{figure}[ht!]
\includegraphics[scale=1]{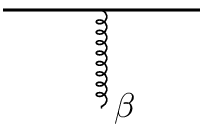}
\includegraphics[scale=1]{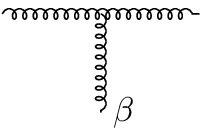}
\includegraphics[scale=1]{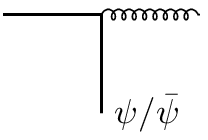}
\includegraphics[scale=1]{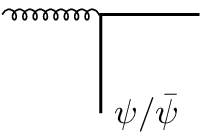}
\caption{Four types of sub-eikonal vertices involved in small-$x$ helicity evolution. Horizontal lines indicate the propagation along the eikonal trajectory of the probe, and vertical lines indicate the interaction with background field generated by the target within the shock-wave. Straight lines indicate quarks or antiquarks, and are either in the fundamental representation or in its conjugate. 
}
\label{FIG:SubEik_int_vertices}
\end{figure}

In order to adapt this method to helicity dependent scattering, let us first introduce the emission operators involved. From \cite{Kovchegov:2015pbl,Kovchegov:2016zex,Kovchegov:2016weo,Kovchegov:2017jxc,Kovchegov:2017lsr,Kovchegov:2018znm} we see that the helicity evolution involves the eikonal interaction vertices along with the sub-eikonal helicity-dependent interaction vertices. Those latter ones are shown in \fig{FIG:SubEik_int_vertices}, where only sub-eikonal contribution is implied in the first two panels from the left. This is denoted by the label $\beta$ on the $t$-channel gluon line coming from the target shock wave (from below). It is defined by \cite{Kovchegov:2017lsr,Kovchegov:2018znm}
\begin{align}\label{beta_def}
\beta (x) = F_{12} (x) = \epsilon^{ij}_\perp \partial_i A_j (x),
\end{align}
where the Latin indices are $i,j = 1,2$, $\epsilon^{ij}_\perp$ is a two-dimensional Levi-Civita symbol, and $F_{12}$ is the gluon field strength tensor. As shown in \cite{Kovchegov:2017lsr,Kovchegov:2018znm}, the transfer of helicity from the target shock wave to the probe in the gluon $t$-channel sector is proportional to this sub-eikonal operator $\beta (x)$. Helicity can be transferred between the target and projectile via an (anti)quark exchange, as shown in the right two panels of \fig{FIG:SubEik_int_vertices}. This means that the sub-eikonal shock wave fields should also include quark and antiquark $\psi (x)$ and ${\bar \psi} (x)$ (in addition to $\beta (x)$). 

The average over the target configuration in helicity-dependent scattering is thus dependent on those additional fields $\beta, \psi, {\bar \psi}$.
Let us extent the weight functional ${\cal W}_Y[\alpha]$ to be also a functional of those fields ${\cal W}_Y[\alpha,\beta,\psi,\bar{\psi}]$. Then the expectation value of a helicity-dependent operator in the longitudinally polarized target state is
\begin{equation}
\label{EQ:Operator_average_helicity}
\langle \hcalo_{pol} \rangle_Y = \int \cald\alpha \cald\beta \cald\psi \cald \bar{\psi} \ \hcalo_{pol} \ {\cal W}_Y[\alpha,\beta,\psi,\bar{\psi}].
\end{equation}
It is understood that ${\cal W}_Y[\alpha,\beta,\psi,\bar{\psi}]$ is also normalized to one (cf. \eq{Wnorm}).

To apply the method described in \cite{Mueller:2001uk,Kovchegov:2012mbw}, we first need to introduce the definition of polarized Wilson lines. Those will allow us to define our test operator $\hcalo_{pol}$. Evolving this operator in rapidity will give us the evolution kernel  $\calk_h$ for small-$x$ helicity evolution,
\begin{align}
\label{EQ:Operator_evo_helicity}
& \langle \hcalo_{pol} \rangle_Y = \langle \hcalo_{pol} \rangle_0 \\ & + \int \cald\alpha \cald\beta \cald\psi \cald \bar{\psi} \int dy \, \left( \calk_h (Y, y) \cdot   \hcalo_{pol} \right) \,{\cal W}_y  . \notag
\end{align}
Note that helicity evolution equations are written as integral equations, without the $\pd_Y$ derivative on the left-hand side, but with an extra rapidity integral on the right \cite{Kovchegov:2015pbl,Kovchegov:2016zex,Kovchegov:2016weo,Kovchegov:2017jxc,Kovchegov:2017lsr,Kovchegov:2018znm} as compared to the unpolarized small-$x$ evolution. This is largely due to the double-logarithmic nature of helicity evolution at leading order, which makes evolution equations much easier to write in the integral form, as will be detailed below. The evolution kernel $\calk_h (Y, y)$ may depend on rapidity, though, as will become apparent later, at the leading order this dependence will come in only as theta-functions constraining the range of the integral over $y$. The inhomogeneous term $\langle \hcalo_{pol} \rangle_0$ in \eq{EQ:Operator_evo_helicity} is given by the initial conditions for the evolution: unlike the unpolarized case, it may have some weak (usually at most linear) dependence on $Y$ \cite{Kovchegov:2015pbl,Kovchegov:2016zex,Kovchegov:2017lsr}. 

One of the main tools of the derivation \cite{Mueller:2001uk,Kovchegov:2012mbw} is the replacement of color generators by functional derivatives using formulas like those in \req{EQ:WilsonLine_Derivative_with respect to_alpha}.
For the flavor-singlet helicity evolution the diagrammatic content is richer than in the unpolarized case, and thus one needs to generalize  \req{EQ:WilsonLine_Derivative_with respect to_alpha} to the case of polarized Wilson lines. This generalization will involve functional derivatives with respect to the the sub-eikonal fields $\beta$, $\psi$, and $\bpsi$ in the kernel $\calk_h  (Y, y)$.

Finally we will need to integrate the functional integrals by parts in \eq{EQ:Operator_evo_helicity}: we compare the result, with
$\calk_h (Y,y)$ acting on the weight functional ${\cal W}_Y$ on the right-hand side of \eqref{EQ:Operator_evo_helicity}, to \eq{EQ:Operator_average_helicity}. 
Arguing that the resulting relations are valid for any helicity-dependent test operator $\hcalo_{pol}$ we should arrive at the helicity-dependent generalization of the JIMWLK equation, 
\begin{align}\label{hJ1}
{\cal W}_Y = {\cal W}_Y^{(0)} + \int dy \ \calk_h (Y, y) \cdot {\cal W}_y .
\end{align}
In obtaining \eq{hJ1} we have also assumed that the kernel $\calk_h (Y,y)$ does not get modified in the integration by parts. This will be demonstrated explicitly below.


\subsection{Functional derivatives and notations}


\paragraph{Polarized Wilson lines.} 
Those are derived in \cite{Kovchegov:2017lsr,Kovchegov:2018znm} for the (anti)quarks and gluons scattering on a polarized target. They consist of light-cone Wilson lines with the sub-eikonal operator insertions shown in \fig{FIG:SubEik_int_vertices}. We will distinguish two types of contributions: (i) sub-eikonal $t$-channel gluon-exchange contributions, involving one insertion of the $\beta$-field, as shown in the left two panels of \fig{FIG:SubEik_int_vertices} for the quark and gluon polarized Wilson lines respectively, and (ii) involving two sub-eikonal $t$-channel quark exchanges, shown in the right two panels of \fig{FIG:SubEik_int_vertices}. The former will be indicated with a superscript $g$ and the latter with a superscript $q$. 

\begin{figure}[ht!]
\includegraphics[scale=1]{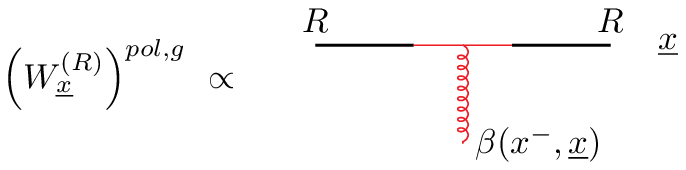}
\caption{Contribution to the polarized Wilson line with a gluon in the t-channel. \ColorOnline}
\label{PolWL-g}
\end{figure}

The contribution (i) involves a single sub-eikonal interaction with the background field $\beta$ that corresponds to the $z$ component of the chromo-magnetic field generated by the target (see the discussion bellow Eq.~(17) in \cite{Kovchegov:2017lsr}). Generalizing the infinite polarized Wilson lines of \cite{Kovchegov:2017lsr,Kovchegov:2018znm} to the case of the finite extent we write (see Eq.~(21) in \cite{Kovchegov:2017lsr} or Eqs.~(44) and (54) in \cite{Kovchegov:2018znm})
\begin{align}
\label{EQ:Def_of_W_pol_g}
& W^{(R) \, pol,g}_\xvec [b^-, a^-] = \eta_R
\frac{ig p_1^+}{s} \, \theta (b^-) \, \theta (-a^-) \nonumber \\
& \times \, \int\limits_{a^-}^{b^-} \text{d}x^-\, W^{(R)}_\xvec [b^- , x^-] \, \beta(x^-,\xvec) \, W^{(R)}_\xvec [x^- , a^- ] 
\end{align}
with 
\begin{align}\label{eta_def}
\eta_R = \delta_{R=F} + \delta_{R=\bar{F}} + 2 \delta_{R=A}. 
\end{align}
(Here $p_1^+$ is the large light-cone momentum of the target proton, $s$ is the center-of-mass energy squared for the projectile-target scattering, $F$ and $\bar{F}$ stand for the fundamental and anti-fundamental representations, while $A$ denotes the adjoint representation.) Analogously to the unpolarized case we define
\begin{align}
W^{(R) \, pol,g}_\xvec \equiv W^{(R) \, pol,g}_\xvec [\infty, -\infty].
\end{align}
Diagrammatically \eq{EQ:Def_of_W_pol_g} can be represented as shown in \fig{PolWL-g}  (up to the prefactor and the integration measure $\text{d}x^-$).

In our diagrammatic representation, thick straight black lines always represent eikonal Wilson lines in a given irrep $R$. Thin red (gray) lines  denote only color tensors. The center part of the \fig{PolWL-g} diagram in red (gray) is thus just the color generator of representation $R$ needed for the background field $\beta$ to be in this representation: $\beta (x) = \beta^a (x) t^a_R$. This notation will soon be useful for visualizing some of the color algebra using diagrams. 

\begin{figure}[ht]
\includegraphics[scale=.89]{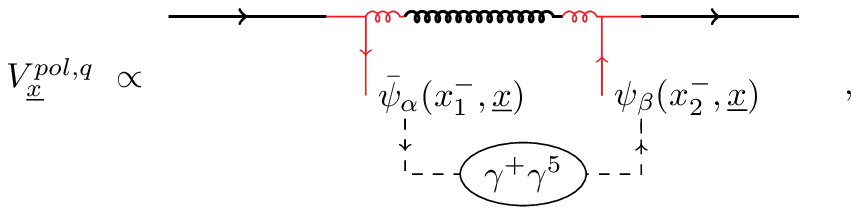}
\includegraphics[scale=.89]{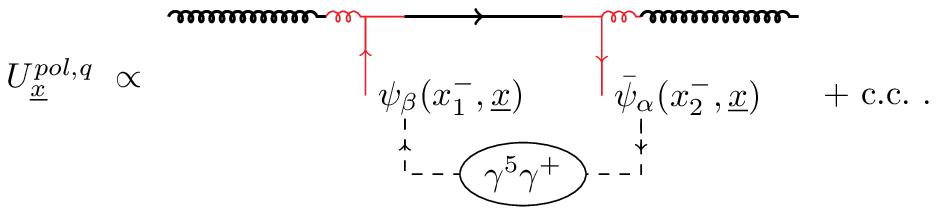}
\caption{Contributions to the fundamental and adjoint polarized Wilson lines from the quark exchanges in the $t$-channel. The dashed lines are used to indicate the contraction over the spinor indices (as opposed to regular straight lines denoting multiplication in color space). \ColorOnline}
\label{FIG:PolWL-q}
\end{figure}

\begin{widetext}
The contribution (ii) involves the interaction with the background fields $\psi$ and $\bar{\psi}$. Since we are only considering the flavor-singlet case, they always contribute in pairs \cite{Kovchegov:2016zex,Kovchegov:2018znm}. 
For the fundamental representation we have (see Eq.~(50) in \cite{Kovchegov:2018znm})
\begin{align}
\label{EQ:V_pol_q}
V_\xvec^{pol,q} [b^-, a^-] = & - \frac{g^2 p_1^+}{s} \, \theta (b^-) \, \theta (-a^-) \\ & \times \, \int \limits_{a^-}^{b^-} \text{d}x_1^-  \int \limits_{x_1^-}^{b^-} \text{d}x_2^- V_\xvec [ b^- , x_2^-] t^b \psi_\beta(x_2^-,\xvec) U_\xvec^{ba} [x_2^- , x_1^- ] \left[ \tfrac{1}{2} \gamma^+ \gamma^5\right] _{\alpha\beta} \bar{\psi}_\alpha(x_1^-,\xvec) t^a V_\xvec [x_1^- , a^-] . \notag
\end{align}
In our notation, $V$ denotes fundamental Wilson lines and fundamental polarized Wilson line operators. 
For the adjoint representation we have (see Eq.~(63) in \cite{Kovchegov:2018znm})
\begin{align}
\label{EQ:U_pol_q}
\left( U_\xvec^{pol,q} [b^-, a^-] \right)^{ab} = & - \frac{g^2 p_1^+}{s} \, \theta (b^-) \, \theta (-a^-) \\ & \times \, \int \limits_{a^-}^{b^-} \text{d}x_1^-  \int \limits_{x_1^-}^{b^-} \text{d}x_2^- U^{aa'} [b^- , x_2^-] \bar{\psi}(x_2^-,\xvec) t^{a'} V_\xvec [x_2^-, x_1^-]\left[ \tfrac{1}{2} \gamma^+ \gamma^5\right] t^{b'} \psi(x_1^-,\xvec) U_\xvec^{b'b} [x_1^- , a^- ]  + \mbox{c.c.} . \notag
\end{align} 
\end{widetext}
The infinite polarized Wilson lines are denoted by
\begin{equation}
V^{pol,q}_\xvec \equiv V^{pol, q}_\xvec [\infty, -\infty], \ \ \ U^{pol,q}_\xvec \equiv U^{pol, q}_\xvec [\infty, -\infty].
\end{equation}
Using the previously defined convention for the diagrammatic representation, the complicated expressions \eqref{EQ:V_pol_q} and \eqref{EQ:U_pol_q} are represented in \fig{FIG:PolWL-q}.

Finally, the net polarized Wilson lines in the fundamental and adjoint representations are given by \cite{Kovchegov:2017lsr,Kovchegov:2018znm}
\begin{subequations}\label{EQ:UV_pol}
\begin{align}
\label{EQ:V_pol}
V_\xvec^{pol} &= V_\xvec^{pol,g} + V_\xvec^{pol,q},  \\
\label{EQ:U_pol}
U_\xvec^{pol} &= U_\xvec^{pol,g} + U_\xvec^{pol,q}  
\end{align}
\end{subequations}
for both infinite and finite longitudinal extent of the lines.  A polarized Wilson line in an arbitrary representation $R$ (including the fundamental and adjoint ones in Eqs.~\eqref{EQ:UV_pol}) is diagrammatically represented by a Wilson line with a gray box, 
\begin{center}
\includegraphics{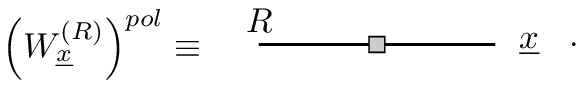}
\end{center}

\smallbreak
\paragraph{Functional derivatives of polarized Wilson lines.}
In order to apply the method presented above in Sections~\ref{Sec:2A} and \ref{Sec:2B} to helicity evolution, we will need relations similar to Eqs.~\eqref{EQ:WilsonLine_Derivative_with respect to_alpha} and \eqref{EQ:WilsonLine_Derivative_with respect to_alpha_2} involving polarized Wilson lines.
Using the definitions \eqref{EQ:V_pol} and \eqref{EQ:U_pol}, one can show in the polarized case that the functional derivative with respect to the eikonal field $\alpha (x)$ reads
\begin{align}
& \frac{\delta\, W^{(R)\,pol,g}_\xvec}{\delta\, \alpha^a(y^-,\yvec)} = ig\, \delta^{(2)}(\yvec - \xvec) 
\label{EQ:W_pol_g_D_alpha} \\
& \times
\begin{cases}
W^{(R)\,pol,g}_\xvec [\infty, y^-] \,t^a_R\, W^{(R)}_\xvec [y^-, -\infty] & \text{for } y^- < 0, \\
W^{(R)}_\xvec  [\infty, y^-] \,t^a_R\, W^{(R)\,pol,g}_\xvec [y^-, -\infty] & \text{for } y^- > 0, 
\end{cases} \nonumber
\end{align}
in an almost direct analogy to \eq{EQ:WilsonLine_Derivative_with respect to_alpha}. Similar to \req{EQ:WilsonLine_Derivative_with respect to_alpha_2}, the two unpolarized Wilson lines in \eq{EQ:W_pol_g_D_alpha} can be replaced by identities in color space as the shock-wave is located at $x^-=0$.

Functional differentiation with respect to $\alpha (x)$ of the polarized Wilson lines with superscript $q$ is a bit more involved. Potential complications may arise from differentiating $U_\xvec^{ba} [x_2^- , x_1^- ]$ in \eq{EQ:V_pol_q} and $V_\xvec [x_2^-, x_1^-]$ in \eq{EQ:U_pol_q}. However, in the leading logarithmic (and double logarithmic) in energy resummations one always has an ordering of light-cone $x^-$-lifetimes, such that the subsequent evolution after each emission is absorbed in a shock wave of a shorter $x^-$-extent than the lifetime of the emitted parton. This is why we always put the unpolarized Wilson lines not crossing the shock wave to unity after differentiation: we simply say that the gluon field in those Wilson lines is zero, $A^+ =0$, outside the shock wave, making them trivial. Following the same logic, is we differentiate the Wilson lines \eqref{EQ:V_pol_q} and \eqref{EQ:U_pol_q} in the $x_1^- < y^- < x_2^-$ region, we have to either put $\psi =0$ or ${\bar \psi} =0$ as being outside the remaining (next-step) shock wave. We would then get zero. Hence, with this logarithmic approximation in mind, we do not differentiate $U_\xvec^{ba} [x_2^- , x_1^- ]$ in \eq{EQ:V_pol_q} and $V_\xvec [x_2^-, x_1^-]$ in \eq{EQ:U_pol_q}.

It is then straightforward to write
\begin{align}
& \frac{\delta\, W^{(R)\,pol,q}_\xvec}{\delta\, \alpha^a(y^-,\yvec)} = ig\, \delta^{(2)}(\yvec - \xvec) 
\label{EQ:W_pol_q_D_alpha} \\
& \times
\begin{cases}
W^{(R)\,pol,q}_\xvec [\infty, y^-]  \,t^a_R\, W^{(R)}_\xvec [y^-, -\infty] & \text{for } y^- < 0, \\
W^{(R)}_\xvec [\infty, y^-] \,t^a_R\, W^{(R)\,pol,q}_\xvec [y^-, -\infty] & \text{for } y^- > 0.
\end{cases} \nonumber
\end{align}

We construct the $\alpha$-derivative of the full polarized Wilson line by summing Eqs.~\eqref{EQ:W_pol_g_D_alpha} and \eqref{EQ:W_pol_q_D_alpha}, obtaining
\begin{align}
\label{EQ:W_pol_D_alpha}
\frac{\delta\, W^{(R)\,pol}_\xvec}{\delta\, \alpha^a(y^-,\yvec)} &= ig \, \delta^{(2)}(\yvec - \xvec)
\begin{cases}
W^{(R)\,pol}_\xvec  \,t^a_R & \text{for } y^- < 0, \\
t^a_R\, W^{(R)\,pol}_\xvec & \text{for } y^- > 0, 
\end{cases}
\end{align}
where the unpolarized Wilson lines that do not cross the shock-wave are again set to identity in the color space.

Moving on to the functional derivative with respect to $\beta (x)$, one first observes that $\beta (x)$ only appears in the contributions with superscript $g$. Using the definition \eqref{EQ:Def_of_W_pol_g}, one finds
\begin{align}\label{beta_der}
\frac{\delta\, W^{(R)\,pol}_\xvec}{\delta\, \beta^a(y^-,\yvec)} =& \, 
\eta_R \frac{ig p_1^+}{s} \ \delta^{(2)}(\yvec-\xvec) \\
&\times \, W^{(R)}_\xvec [\infty , y^-] \,t^a_R\,  W^{(R)}_\xvec [y^- , -\infty ] . \nonumber
\end{align}
Notice the presence of the $\eta_R$ factor on the right-hand side. This factor is not re-absorbed into the definition of a polarized Wilson line since the right-hand side of \eqref{beta_der} does not contain any, unlike the above case of the functional derivatives with respect to $\alpha (x)$.

We next need to derive the functional derivatives with respect to the fermion background fields. For the flavor-singlet evolution case, we are interested in the operator
\begin{equation}
\label{EQ:Diff_op_psi_bpsi}
\frac{\delta^2}{\delta\, \bar{\psi}_{\alpha,i}(y_1^-,\yvec_1) \ \delta\, \psi_{\beta,j}(y_2^-,\yvec_2)}
\end{equation}
where $\{\alpha,\beta\}$ subscripts indicate the spinor space and $\{i,j\}$ are color indices. To pick up the helicity-dependent part we will need to contract the remaining tensor with  the Brodsky-Lepage spinors $\chi$ \cite{Lepage:1980fj}
\begin{equation}\label{hel_proj}
\tfrac{1}{\sqrt{2}}\,\sum_{\xi=\pm 1} \xi\  \chi_\alpha(\xi) \chi_\beta(\xi) = \tfrac{1}{2}(\gamma^5\gamma^-)_{\beta\alpha}
\end{equation}
which would remove the $\tfrac{1}{2} \gamma^+\gamma^5$ matrix in Eqs.~\eqref{EQ:V_pol_q} and \eqref{EQ:U_pol_q}, since
$\tfrac{1}{2}(\gamma^5\gamma^-)_{\beta\alpha} \cdot  \tfrac{1}{2}(\gamma^+\gamma^5)_{\alpha\beta} = 1$.

The functional derivatives of interest are as follows.
\begin{widetext}
\noindent Acting with \req{EQ:Diff_op_psi_bpsi} on the fundamental polarized Wilson line gives
\begin{align}
\frac{\delta^2\ (V^{pol,q}_\xvec)_{j'i'}}{\delta\, \bar{\psi}_{\alpha,i}(y_1^-,\yvec_1) \ \delta\, \psi_{\beta,j}(y_2^-,\yvec_2)} =& \ 
\delta^{(2)}(\yvec_1-\xvec) \, \delta^{(2)}(\yvec_2-\xvec) \left(ig\sqrt{p_1^+/s}\right)^2 \left(\frac{\gamma^+\gamma^5}{2}\right)_{\alpha\beta} \nonumber\\ \label{EQ:V_pol_d_bpsi_psi}
&\times \, \theta(y_2^-) \, \theta ( - y_1^-)\ 
\left(V_\xvec [\infty , y_2^-] \, t^b\right)_{j'j} 
U_\xvec^{ba} [y_2^- , y_1^-]
\left(t^a V_\xvec [y_1^- , -\infty] \right)_{ii'} .
\end{align}
The right-hand side is a rank-four tensor in color space valued\footnote{The (complex) vector spaces $V$ and its dual $\bar{V}$ are associated with the fundamental irrep and its conjugate, and the (real) vector space $A$ is associated to the adjoint irrep.} in $V^{\otimes 2}\otimes \bar{V}^{\otimes 2}$ and, as expected, a rank-two tensor in spinor space.
Acting with \req{EQ:Diff_op_psi_bpsi} on the adjoint polarized Wilson line yields
\begin{align}
\frac{\delta^2 \ (U^{pol,q}_\xvec)^{ab}}{\delta\, \bar{\psi}_{\alpha,i}(y_1^-,\yvec_1) \ \delta\, \psi_{\beta,j}(y_2^-,\yvec_2)} = \ & - \
\delta^{(2)}(\yvec_1 -\xvec) \, \delta^{(2)}(\yvec_2 -\xvec) \left(ig\sqrt{p_1^+/s}\right)^2
\left(\frac{\gamma^+\gamma^5}{2}\right)_{\alpha\beta} \nonumber\\ \label{EQ:U_pol_d_bpsi_psi}
&\times \Bigg[ \theta(y_1^- ) \, \theta (  - y^-_2) \
U^{aa'} _\xvec [\infty , y_1^-] \ 
\left(t^{a'} \, V_\xvec [y_1^- , y_2^-]  \, t^{b'} \right)_{ij}
U^{b'b} _\xvec [y_2^- , -\infty] \notag \\ & + \theta(y_2^- ) \, \theta (  - y^-_1) \ U^{aa'} _\xvec [\infty , y_2^-] \ 
\left(t^{b'} \, V_\xvec [y_1^- , y_2^-]  \, t^{a'} \right)_{ij}
U^{b'b} _\xvec [y_1^- , -\infty] \Bigg] . 
\end{align}
\end{widetext}
This time the right-hand side is valued in the color space $A^{\otimes 2}\otimes V \otimes \bar{V}$. 
Notice the relative minus sign in the right-hand side of \req{EQ:U_pol_d_bpsi_psi} compared to \req{EQ:V_pol_d_bpsi_psi}. It follows from the fact that the fields $\psi$ and $\bar{\psi}$ are Grassmann-valued. Due to the increasing number of indices, it is also useful to visualize Eqs.~\eqref{EQ:V_pol_d_bpsi_psi} and \eqref{EQ:U_pol_d_bpsi_psi} in terms of diagrams in color space (dropping the symbols for the fermion fields to indicate functional differentiation):
\begin{align}
\mbox{\req{EQ:V_pol_d_bpsi_psi}} \propto \parbox{68mm}{\includegraphics[width=68mm]{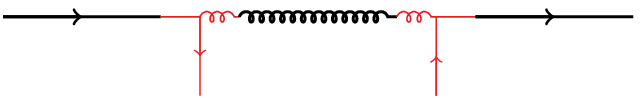}} , \notag \\
\mbox{\req{EQ:U_pol_d_bpsi_psi}} \propto \parbox{68mm}{\includegraphics[width=68mm]{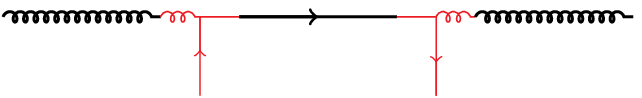}} . \notag
\end{align}
Just as before, the unpolarized Wilson lines in Eqs.~\eqref{EQ:V_pol_d_bpsi_psi} and \eqref{EQ:U_pol_d_bpsi_psi} not crossing the shock wave have to be put to unity in the end.

We are now in a position to construct the evolution kernel for a polarized test operator. However, before doing so let us introduce some shorthand notation for the functional derivatives.

\smallbreak
\paragraph{Notations.} Denote an abbreviated notation for the functional derivative with respect to the field $\alpha (x)$ as
\begin{equation}
D^+_{{\un y}, a, \lessgtr} \equiv (ig)^{-1}\  \frac{\delta}{\delta\, \alpha^a(y^-\lessgtr 0,\yvec)} .
\end{equation}
For the $\beta(x)$ field we write
\begin{equation}
D^\perp_{{\un y}, a,\lessgtr} \equiv \left(\frac{ig\, p^+_1}{s}\right)^{-1}
\frac{\delta}{\delta\, \beta^a(y^-\lessgtr 0,\yvec)} .
\end{equation}
For the quark fields we have 
\begin{subequations}
\begin{align}
D^{\psi}_{{\un y},\alpha,i,\lessgtr} \equiv \left(ig\sqrt{p_1^+/s}\right)^{-1}
\frac{\delta}{\delta\, \psi_{\alpha,i}(y^-\lessgtr 0,\yvec)},  \\
D^{\bar{\psi}}_{{\un y}, \alpha,i,\lessgtr} \equiv \left(ig\sqrt{p_1^+/s}\right)^{-1}
\frac{\delta}{\delta\, \bar{\psi}_{\alpha,i}(y^-\lessgtr 0,\yvec)}.
\end{align}
\end{subequations}


\section{Evolution of a polarized test operator}
\label{Sec:3}


\subsection{Evolution kernel for a test operator}\label{Sec:3a}

Our test operator consists of two Wilson lines in an arbitrary color representations $\{R_1,R_0\} = \{F,A\}$ or $\{R_1,R_0\} = \{F,\bar{F},A\}$ at the transverse positions $\xvec_1$ and $\xvec_0$.\footnote{There are two possible conventions for the color representations one can follow here: one may use the standard convention where the anti-quarks are described by a line in the fundamental representation $F$ with the arrow pointing to the left (back in time) in the Feynman diagrams for the scattering amplitude, corresponding to the conjugate Wilson lines $V^\dagger_{\un x}$. Alternatively, one could explicitly introduce the conjugate representation $\bar F$ with the generators $t^a_{\bar F} = - (t^a_F)^T$, and keep all Wilson lines future-pointing. We will try to keep our notation applicable to both conventions.  } The polarized line is chosen at $\xvec_1$ for the graphical representation, but this consideration will not impact the derivation of the evolution kernel in the end. We write
\begin{equation}
\label{EQ:Test_operator_pol}
\hcalo_{\ov,\zv} = s \, \left( W^{(R_0) \, \dagger}_\zv \otimes W^{(R_1)\, pol}_\ov + \mbox{c.c.}. \right)
\end{equation}
The factor of the center-of-mass energy squared $s$ in \eqref{EQ:Test_operator_pol} cancels the sub-eikonal factor of $1/s$ in the polarized Wilson lines from Eqs.~\eqref{EQ:Def_of_W_pol_g}, \eqref{EQ:V_pol_q}, and \eqref{EQ:U_pol_q}. The factor of $p_1^+$ in those equations cancels the sub-eikonal factors in the fields $\beta(x) \sim 1/p_1^+$ and $\psi \sim {\bar \psi} \sim 1/\sqrt{p_1^+}$ \cite{Kovchegov:2017lsr,Kovchegov:2018znm}. Therefore, the operator \eqref{EQ:Test_operator_pol} has no explicit energy dependence. The energy dependence of its expectation value $\langle \hcalo_{\ov,\zv} \rangle$ will be generated by the small-$x$ evolution we are about to construct. It can be thought of as rapidity dependence, which would come in as the dependence on the slope (in relation to the light cone) of the Wilson lines comprising this operator \cite{Balitsky:2008zza,Balitsky:2015qba}. In JIMWLK formalism the energy dependence comes in through the weight functional ${\cal W}_Y$.

The complex conjugate term is added in \eq{EQ:Test_operator_pol} to single out the flavor-singlet contribution \cite{Kovchegov:2016zex,Kovchegov:2018znm}. Thus the resulting helicity JIMWLK kernel we are about to construct would apply only to evolution of such flavor-singlet helicity operators. Generalization of our results to the flavor non-singlet case is left for future work. 

\begin{figure}[h]
\includegraphics[scale=1]{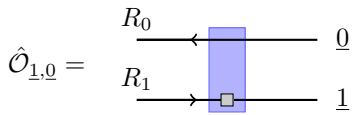}
\caption{Test operator $\hcalo_{\ov,\zv}$ that will be evolved in rapidity. The shock wave is represented in blue (dark shade of gray) and the polarized Wilson line is indicated by the light gray box. \ColorOnline}
\label{FIG:Test_operator}
\end{figure}

The operator \eqref{EQ:Test_operator_pol} (without the complex conjugate term) is illustrated in \fig{FIG:Test_operator}.

\smallbreak
\paragraph{Lifetime and longitudinal momentum ordering.}

Helicity evolution at the leading order resums two logarithms of Bjorken $x$ per each power of the strong coupling $\as$ \cite{Kovchegov:2015pbl,Kovchegov:2016zex,Kovchegov:2017lsr,Kovchegov:2018znm}. This type of resummation is known as the double-logarithmic approximation (DLA) and results in $\as \, \ln^2 (1/x)$ being the resummation parameter \cite{Kirschner:1983di, Kirschner:1985cb,Kirschner:1994vc,Kirschner:1994rq,Bartels:1996wc,Bartels:1995iu,Griffiths:1999dj,Itakura:2003jp,Bartels:2003dj}. Note that this situation is different from the unpolarized small-$x$ evolution \cite{Kuraev:1977fs,Balitsky:1978ic,Balitsky:1995ub,Balitsky:1998ya,Kovchegov:1999yj,Kovchegov:1999ua,Jalilian-Marian:1997dw,Jalilian-Marian:1997gr,Weigert:2000gi,Iancu:2001ad,Iancu:2000hn,Ferreiro:2001qy}, where leading-order resummation parameter is $\as \, \ln (1/x)$, that is, it contains only one logarithm of $x$.

Double logarithmic resummation requires quark and gluon emissions to be ordered in two kinematic parameters. For helicity evolution the two parameters are the light-cone momentum fraction $z$ and the lifetime of the fluctuation in the $x^-$ direction. The two Wilson lines in our test operator \eqref{EQ:Test_operator_pol} may represent a quark and an anti-quark propagators carrying light-cone momentum fractions $z_0$ and $z_1$ of the net projectile momentum respectively. That is, the two lines carry momenta $k_0^- = z_0 \, p^-$ and $k_1^- = z_1 \, p^-$, where $p^-$ is the large momentum of some projectile that gave rise to the quark and anti-quark in our test operator. Alternatively, one can say that the center-of-mass energies squared of each of these lines scattering on the shock-wave target are $z_0 s$ and $z_1 s$. In DLA, the expectation value of the test operator \eqref{EQ:Test_operator_pol} is a function of the smallest of the two $z$'s,  $\langle \hcalo_{\ov,\zv} \rangle = \langle \hcalo_{\ov,\zv} \rangle (\min \{z_0, z_1 \} \, s)$ \cite{Kovchegov:2015pbl}. As we pointed out above, this dependence is not explicit in \eq{EQ:Test_operator_pol} and would come in through small-$x$ evolution. 

In helicity evolution, one power of $\ln (1/x)$ (per one $\as$) arises from the ordering of the light-cone momentum fractions $z$ of the emitted partons,
\begin{align}\label{ordering1}
z_2 \gg z_3 \gg z_4 \gg \ldots ,  
\end{align}
where the number in the subscript counts successive emissions. The other $\ln (1/x)$ arises from the transverse coordinate integration, after imposing the $x^-$-lifetime ordering. For an emission of a small-$z$ gluon or quark with momentum $k^\mu$ (such that $k^- = z \, p^-$), the $x^-$-lifetime is $2 k^-/{\un k}^2 \sim z \, {\un X}^2$, where $\un X$ is the transverse space vector Fourier-conjugate to $\un k$. The lifetime ordering giving rise to the second power of $\ln (1/x)$ is  (see Sec.~IV of \cite{Kovchegov:2015pbl})
\begin{align}\label{ordering2}
z_2 \, {\un X}_{2}^2 \gg z_3 \, {\un X}_{3}^2 \gg z_4 \, {\un X}_{4}^2 \gg  \ldots .  
\end{align}
(We will explicitly define the vectors ${\un X}_n$ below.) 

The fact that helicity evolution in DLA arises from simultaneously imposing the light-cone momentum fraction ordering \eqref{ordering1} and the lifetime ordering \eqref{ordering2} is explained in detail in Appendix~\ref{app}, which presents a new important cross check of the need for both orderings. 

To construct helicity JIMWLK, we will work with the light-cone momentum fractions $z$ and $z'$ for the operators before and after one step of evolution, such that $Y \sim \ln (z s)$ and $y \sim \ln (z' s)$ in Eqs.~\eqref{EQ:Operator_evo_helicity} and \eqref{hJ1}. Since $z \gg z'$ we have $Y \gg y$, which means that the $y$-integral in \eqref{EQ:Operator_evo_helicity} is cut off by $Y$ from above. This easily imposes the light-cone momentum fraction ordering \eqref{ordering1} in our evolution. We will also need to impose the $x^-$-lifetime ordering \eqref{ordering2}. As we will shortly see, this implies inserting cutoffs on the transverse coordinate integrations \cite{Kovchegov:2015pbl,Kovchegov:2016zex,Kovchegov:2017lsr,Kovchegov:2018znm} which depend on $z'$ and on the lifetime at the previous step of the evolution. This means the cutoffs will depend on the transverse distance(s) between some earlier emitted partons with respect to their ``parent" partons, that is, the partons off of which those earlier partons were emitted.  It appears that to impose lifetime ordering we need to generalize \eq{EQ:Operator_evo_helicity} to have the integral kernel explicitly depend on the transverse distances determining $x^-$-lifetimes at the previous step of the evolution.

\begin{figure}
\includegraphics[scale=1]{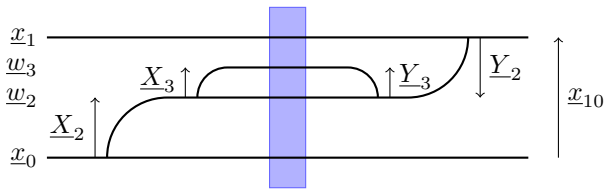}
\caption{A possible contribution to the evolution involving emission of two partons. Represented lines can be polarized or unpolarized and can also be in any color representation allowed by Feynman rules. The vertices involved can be eikonal or sub-eikonal.
This specific diagram corresponds to $\{\xvec_{(1)} = \xvec_0,\, \xvec_{(2)}=\wvec_2\}$ on the left of the shock wave and $\{\yvec_{(1)} = \xvec_1,\, \yvec_{(2)} =\wvec_2\}$ on the right of the shock wave. \ColorOnline
}
\label{FIG:Ordering}
\end{figure}

Let us denote by $\wvec_n$ the transverse position of the $n$th parton emitted with the fraction of the projectile's light-cone ``minus'' momentum $z_n$ with $n = 2, 3, \ldots$.\footnote{In cases where only a few partons are involved, we will also use a simpler notation with $z'=z_2$, $z''=z_3$.}
As was the case with the JIMWLK evolution, all relevant diagrams are generated by the use of the Leibniz rule for the functional derivatives acting on the test operator.
When acting on the left (resp. right) of the shock wave, let us denote the corresponding coordinate in the functional derivative by $\xvec_{(n)}$ (resp. $\yvec_{(n)}$).
Those dummy coordinates should be distinguished from $\xvec_0$ or $\xvec_1$ corresponding to the positions of the Wilson lines in the test operator. It is useful to define
\begin{equation}\label{XYnot}
\Xvec_{n} = \xvec_{(n)} - \wvec_n \quad \text{and} \quad \Yvec_{n} = \yvec_{(n)} - \wvec_n.
\end{equation}
Consider the diagram shown in \fig{FIG:Ordering}, which contains two additional partons generated by the DLA evolution compared to the test operator defined in \req{EQ:Test_operator_pol} or \fig{FIG:Test_operator}.
Imposing DLA orderings \eqref{ordering1} and \eqref{ordering2} in this diagram, with the lifetime ordering applied separately to the left and to the right of the shock wave \cite{Kovchegov:2015pbl}, we arrive at
\begin{subequations}
\begin{align}
& z \gg z_2 \gg z_3, \\
& z\, x_{10}^2 \gg  z_2 \, (\wvec_2 - \xvec_0)^2  \gg  z_3 \, (\wvec_3 - \wvec_2)^2, \\
& z\, x_{10}^2 \gg  z_2\, (\wvec_2 - \xvec_1)^2  \gg  z_2\, (\wvec_3 - \wvec_2)^2,
\end{align}
\end{subequations}
where $z \equiv \min \{z_0, z_1 \}$ and $x_{01} \equiv |{\un x}_0 - {\un x}_1 |$.
Introducing the notations defined in \eqref{XYnot} we easily generalize this to the case of an arbitrary DLA diagram with two extra partons, \begin{subequations}
\begin{align}
& z \gg z_2 \gg z_3, \\
& z\, x_{10}^2 \gg  z_2 \, X_2^2 \gg  z_3 \, X_3^2, \label{lt_ord2} \\
& z\, x_{10}^2 \gg z_2 \, Y_2 ^2  \gg  z_3 \, Y_3^2.
\end{align}
\end{subequations}

It is now straightforward to write down the ordering conditions for the $(n+1)$st emission, 
\begin{subequations}\label{EQ:Ordering}
\begin{align}
z_n & \, \gg z_{n+1}, \label{EQ:Ordering-a}\\
z_n X_n^2 & \, \gg  z_{n+1} X_{n+1}^2, \label{EQ:Ordering-b}\\
z_n Y_n^2 & \, \gg z_{n+1} Y_{n+1}^2. \label{EQ:Ordering-c}
\end{align}
\end{subequations}
We arrive at the following physical picture: as we increase the available rapidity interval, extra soft partons are emitted.
Those are generated by acting with the evolution kernel $\calk_h$ on the previous state of the projectile. Because we require the longitudinal momentum ordering \eqref{EQ:Ordering-a}, the integral evolution kernel $\calk_h$ depends on the longitudinal fraction of the new parton $z_{n+1}$ but also on the longitudinal fraction of the previously emitted parton $z_{n}$ as shown in \req{hJ1} via the kernel's dependence on rapidities $y$ and $Y$.
In order to also satisfy the lifetime ordering in both Eqs.~\eqref{EQ:Ordering-b} and \eqref{EQ:Ordering-c}, the evolution kernel should also depend on the transverse separations $X_{n+1}, X_n$ and $Y_{n+1}, Y_n$. In other words, due to Eqs.~\eqref{EQ:Ordering-b} and \eqref{EQ:Ordering-c}, the integration range of the $(n+1)$st (``daughter") parton's transverse position is limited not just by the position(s) of the (``parent") parton(s) off of which the $(n+1)$st parton was emitted, but also by the position(s) of some earlier parton(s) relative to their respective ``parent" parton(s), $\{X_n, Y_n\}$, as prescribed by the lifetime-ordering condition.  This is the essential difference of the DLA evolution compared to the unpolarized evolution: while the standard JIMWLK kernel \eqref{EQ:JIMWLK_Kernel} depends on the positions $\un x$ and $\un y$ of the ``parent" partons relative to the daughter parton $\un w$, it is independent of the locations of the partons emitted before the ``parents". 

We see that for helicity evolution we need to bring into the evolution kernel the dependence on the distances between the partons emitted in the previous step of our evolution and their respective ``parent" partons. To do this we introduce an abbreviated notation
\begin{align}\label{tau_def}
\tau \equiv \{ z, z\, X^2, z\, Y^2\}, \ \ \ \tau' \equiv \{ z', z'\, X'^2, z'\, Y'^2\}, 
\end{align}
and rewrite \eq{EQ:Operator_evo_helicity} as
\begin{align}
\label{EQ:Operator_evo_helicity1}
& \langle \hcalo_{pol} \rangle_\tau = \langle \hcalo_{pol} \rangle_0 \\ & + \int \cald\alpha \cald\beta \cald\psi \cald \bar{\psi} \int d^3 \tau' \,  \left( \calk_h \left[ \tau, \tau' \right] \cdot   \hcalo_{pol} \right) \,{\cal W}_{\tau'}  , \notag
\end{align}
where 
\begin{align}\label{d3t}
d^3\tau' \equiv \frac{d z'}{z'} \, \dd X' \dd Y' .
\end{align}
(Note that in DLA the integrals over the angles of ${\un X}'$ and ${\un Y}'$ are trivial: this is why we write $d^3\tau'$ instead of $d^5\tau'$ in \eq{d3t}.  We cannot integrate these angles out from the start though, due to the structure of the kernel $\calk_h$, as we will see below.)

Integration by parts in \eq{EQ:Operator_evo_helicity1} should give us the following helicity-dependent generalization of the JIMWLK evolution (cf. \req{hJ1}):
\begin{equation}
\label{EQ:Evolution}
\calw_{\tau} = \calw^{(0)}_\tau + 
\int d^3 \tau' \, \calk_h \left[ \tau, \tau' \right] \, \calw_{\tau'}.
\end{equation}
The helicity evolution kernel should contain $\theta$-functions imposing the constraints \eqref{EQ:Ordering}. We include those by introducing another abbreviation,
\begin{align}\label{theta3}
 \theta^{(3)}(\tau-\tau') \equiv  \theta(z-z') \, \theta(z \, X^2 - z' \, X'^2) \\ \times \, \theta(z\, Y^2 - z' \, Y'^2) , \notag
\end{align}
with $\theta^{(3)}(\tau-\tau')$ included in all the contributions to $\calk_h$ we consider below. 

Note that for the first step of the evolution of our test operator \eqref{EQ:Test_operator_pol} we have $X_1=Y_1 = x_{10}$. However, as can be seen from the example in \fig{FIG:Ordering}, for subsequent evolution $X_2 \neq Y_2$, and these two transverse distances are different in general. Furthermore, as discussed in detail in \cite{Kovchegov:2015pbl,Kovchegov:2016zex}, there are situations where, say, in the large-$N_c$ limit, evolution of a color-dipole version of the test operator \eqref{EQ:Test_operator_pol} may depend on the size of the ``neighbor" (or ``sister") dipole, which was created by the helicity evolution at the same step as the dipole $10$, but which may have a smaller transverse size, and, hence, dominates the lifetime bound with a shorter lifetime. In the example of \fig{FIG:Ordering}, the large-$N_c$ evolution in dipole $21$ may depend on the size $X_2$ of the dipole $20$ due to the lifetime ordering condition \eqref{lt_ord2}. It can be shown that this situation is accounted for by the $\tau$-notation defined here. In \cite{Kovchegov:2015pbl,Kovchegov:2016zex} such considerations led to the introduction of the ``neighbor" dipole amplitude $\Gamma$ in the large-$N_c$ and large-$N_c \, \& \, N_f$ evolution.

\smallbreak
\paragraph{Eikonal emissions.}
The JIMWLK evolution kernel already includes all the possible diagrams with eikonal emissions. To picture those, consider the diagrams in  \fig{FIG:Evolution_diagrams} and replace the unpolarized Wilson line at $\xvec_1$ by a polarized one (see also \cite{Kovchegov:2015pbl,Kovchegov:2017lsr,Kovchegov:2018znm}). There are two extra considerations to include: (i) One should transform the evolution kernel into an integral kernel, as shown in \eq{EQ:Operator_evo_helicity} or, more precisely, in \eq{EQ:Operator_evo_helicity1}. (ii) The form used in \cite{Mueller:2001uk,Kovchegov:2012mbw} or \req{EQ:JIMWLK_Kernel} for the evolution kernel is no longer obtainable for helicity evolution.

To transform the JIMWLK kernel from a differential to an integral form as per our consideration (i), we need to replace the derivative $\pd_Y$ in \eq{Wevol} by the integral $dy = \text{d}z'/z'$, which is a part of $d^3\tau'$ in \eq{d3t}.  
The origin and structure of the $dy = \text{d}z'/z'$ integral is explained in the discussion of Sec.~IV in \cite{Kovchegov:2015pbl} (see also Eq.~(69) of \cite{Kovchegov:2017lsr}). The integrals $ \dd X' \dd Y'$  in \eq{d3t} are a part of the standard JIMWLK kernel: for the reasons explained above, we will now keep them outside the kernel. 

The reason for the consideration (ii) is that we cannot ``slide'' the color generator across the shock-wave along the polarized Wilson line at $\xvec_1$ by simply adding an adjoint Wilson line at $\xvec_1$. Namely, while the following holds,
\begin{equation}
\label{EQ:Generator_slide_eik}
\left(W^{(R)}_\xvec t^a_R\right) = \left(t^b_R W^{(R)}_\xvec\right) U^{ba}_\xvec ,
\end{equation}
the analogous relation in the polarized case does not,
\begin{equation}
\left(W^{(R),pol}_\xvec t^a_R\right) \neq \left(t^b_R W^{(R),pol}_\xvec\right) U^{ba}_\xvec .
\end{equation}
We can resolve this problem by employing the form of the JIMWLK kernel used in \cite{Kovner:2005jc,Kovner:2005aq}.

\begin{widetext}
\noindent For our test operator \eqref{EQ:Test_operator_pol}, setting $\tau = \{ z, z\, x_{10}^2, z\, x_{10}^2\}$ and $\tau' = \{ z', z'\, X'^2, z'\, Y'^2\}$ with ${\un X}' = \xvec-\wvec$ and ${\un Y}' = \yvec-\wvec$, we rewrite the JIMWLK kernel from \cite{Kovner:2005jc,Kovner:2005aq} as
\begin{align} \label{Keik1}
\calk^{eik}[\tau, \tau'] \equiv \frac{\alpha_s}{\pi^2}  \int \dd w_\perp \frac{{\un X}' \cdot {\un Y}'}{X'^2 \, Y'^2}\ 
\theta^{(3)}(\tau-\tau') \, \theta \left( z' - \tfrac{\Lambda^2}{s} \right) \, \theta \left( X'^2 - \tfrac{1}{z' \, s} \right) \, \theta \left( Y'^2 - \tfrac{1}{z' \, s} \right)  \\ \times \, \left\{ U_\wvec^{ba} \, D^+_{\xvec, a, <} \, D^+_{\yvec, b, >} - \tfrac{1}{2} \left( D^+_{\xvec, a, <} \, D^+_{\yvec, a, <} + D^+_{\xvec, a, >} \, D^+_{\yvec, a, >} \right) \right\}  . \notag 
\end{align}
\end{widetext}
The terms in the kernel \eqref{Keik1} directly correspond to the Feynman diagrams for the eikonal gluon emission and absorption. 
In writing \eq{Keik1} we have augmented the original kernel of \cite{Kovner:2005jc,Kovner:2005aq} by including six $\theta$-functions: three in the $\theta^{(3)}$ defined by \eq{theta3} and another three explicitly shown in \eq{Keik1}. Two of the three $\theta$-functions in \eq{theta3} impose the lifetime ordering in the $x^-$-direction as discussed above (see also Sec.~IV in \cite{Kovchegov:2015pbl}). 
The remaining $\theta$-function in \eq{theta3} imposes the light-cone momentum fraction ordering on the $z'$ integral in the way usual for the leading-logarithmic unpolarized JIMWLK or BK evolution, when written in an integral form. The $z'$-integral also needs a lower cutoff: this is included in \eq{Keik1} with the help of $\theta \left( z' - \tfrac{\Lambda^2}{s} \right)$, where $\Lambda$ is an infrared cutoff momentum scale, such that $\Lambda^2 /s$ is the lowest possible value of any $z_n$ in the problem. Finally, the integrals over  $X'$ and $Y'$ need short-distance (ultraviolet) cutoffs: the shortest available distance squared is the inverse of the available center-of-mass energy squared $z' \, s$. This leads to the last two  $\theta$-functions in \eq{Keik1}.

Only the curly brackets in \eq{Keik1} will change for the other contributions to the helicity evolution kernel $\calk_h$. Anticipating this we will, for brevity, denote everything outside of the curly brackets by $(x\cdot y)$. Further, suppressing the color indices and the position-dependent subscripts in the functional derivatives, we write \eq{Keik1} in the following abbreviated form:
\begin{equation}\label{Keik2}
\calk^{eik} \! \equiv (x\cdot y) \left\{ U_\wvec D^+_<D^+_> - \tfrac{1}{2} ( D^+_< D^+_< + D^+_>D^+_> ) \right\} \! .
\end{equation}
One can easily convince oneself that the above curly bracket matches \req{EQ:JIMWLK_Kernel} by using \req{EQ:Generator_slide_eik}.

While \eq{Keik1} is written for $\tau = \{ z, z\, x_{10}^2, z\, x_{10}^2\}$ defined specifically for the test operator \eqref{EQ:Test_operator_pol}, it is clear that it can be generalized for an arbitrary $\tau$ defined in \eq{tau_def}. Note also that the transverse coordinate integrals one obtains by inserting the kernel \eqref{Keik1} into \eq{EQ:Operator_evo_helicity1} are not logarithmic yet: one would further need to simplify those integrals by identifying the logarithmic region(s) to obtain the DLA contribution along the lines of \cite{Kovchegov:2015pbl,Kovchegov:2017lsr,Kovchegov:2018znm}. This observation will also apply to other contributions to $\calk_h$ we will derive below. 

Finally, let us point out that the lifetime-ordering $\theta$-functions along with $\theta \left( X'^2 - \tfrac{1}{z' \, s} \right) \, \theta \left( Y'^2 - \tfrac{1}{z' \, s} \right)$ do not change the fact that the eikonal kernel \eqref{Keik1} is the standard leading-order JIMWLK kernel. The reasoning behind this observation is that for unpolarized small-$x$ evolution the transverse position integrals are convergent both in the infrared (IR) and in the ultraviolet (UV). In the leading-logarithmic in $\ln (1/x)$ approximation (LLA), the lifetime-ordering $\theta$-functions only provide a cutoff in the deep IR, while $\theta \left( X'^2 - \tfrac{1}{z' \, s} \right) \, \theta \left( Y'^2 - \tfrac{1}{z' \, s} \right)$ provide a cutoff in deep UV: since the integrals in the unpolarized evolution are UV and IR convergent, such cutoffs can be safely neglected.

\begin{figure}[htp]
\includegraphics[scale=.95]{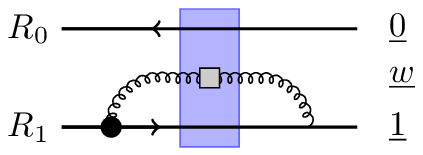}
\hfill
\includegraphics[scale=.95]{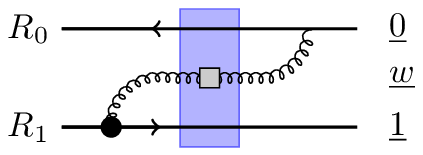}
\includegraphics[scale=.95]{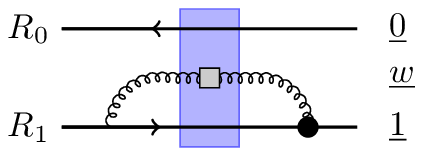}
\hfill
\includegraphics[scale=.95]{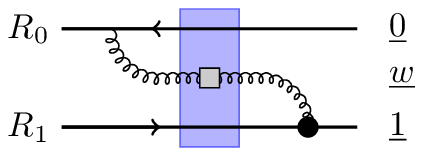}
\caption{The four contributions involving the $\beta$ field derivative after one step of rapidity evolution for the test operator given in \fig{FIG:Test_operator} and in \req{EQ:Test_operator_pol}. The bullet indicates the sub-eikonal emission. \ColorOnline }
\label{FIG:Evolution_beta_field}
\end{figure}

\smallbreak
\paragraph{Sub-eikonal emissions: the $\beta$-field.}
Sub-eikonal diagrams involving the $\beta$ field derivative are given in \fig{FIG:Evolution_beta_field}. The black bullet, indicating a $\beta$-field insertion, appears only on the lower line, since it is the only polarized one and thus $\beta$-dependent. (We are not illustrating the c.c. part of the test operator \eqref{EQ:Test_operator_pol}.)
Those diagrams are generated by the following kernel
\begin{equation}\label{Kbeta}
\calk^{\beta} \equiv (x\cdot y) \left\{ \tfrac{1}{2} U^{pol}_\wvec ( D^+_< D^\perp_> + D^\perp_< D^+_> ) \right\}
\end{equation}
acting on the test operator \eqref{EQ:Test_operator_pol}, as can be seen by comparing with the calculations carried out in \cite{Kovchegov:2018znm}. We employ the abbreviated notation introduced in \eq{Keik2}. 
It is understood that the $D^\perp$ derivatives in $K^{\beta}$ only act on $W^{(R_1),pol}_\ov$ from \fig{FIG:Evolution_beta_field}, since this is the only $\beta$-dependent part of the operator. However, $D^\perp$ can also be allowed to act on $W^{(R_0) \, \dagger}_\zv$ without changing anything, since it is an unpolarized Wilson line that does not depend on the sub-eikonal background field $\beta$. This way $\calk^\beta$ is symmetric under the $x \leftrightarrow y$ interchange, that is, $\calk^\beta$ is now independent of the arbitrary choice of the polarized Wilson line being at the position $\xvec_1$.

The contributions $\calk^{eik}$ and $\calk^{\beta}$ are the only two relevant ones for the helicity evolution in the large-$N_c$ limit. They reproduce the helicity evolution equations derived in \cite{Kovchegov:2017jxc,Kovchegov:2017lsr,Kovchegov:2018znm} at large-$N_c$ while putting $N_f=0$.

\begin{figure}[ht!]
\includegraphics[scale=1]{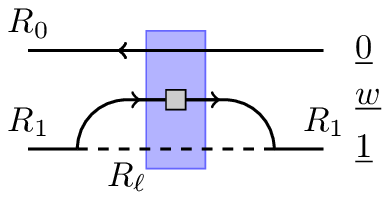}
\hfill
\includegraphics[scale=1]{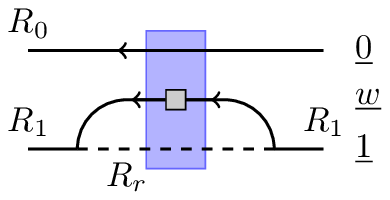}
\caption{The two relevant contributions to the helicity evolution involving the derivatives of the $\psi$ and $\bar{\psi}$ fields.
The dashed line represent the propagation of a Wilson line across the shock-wave in irrep $R_l$ (left) or $R_r$ (right).
Both $R_\ell$ and $R_r$ depends on the irrep $R_1$ as clarified in the text. \ColorOnline }
\label{FIG:Evolution_psi_field}
\end{figure}

\smallbreak
\paragraph{Sub-eikonal emissions: $\psi$ and $\bar{\psi}$  fields.}
The sub-eikonal contribution involving derivatives of the two fields $\psi$ and $\bar{\psi}$ are pictured in \fig{FIG:Evolution_psi_field}. The Wilson line crossing the shock-wave at the transverse position $\xvec_1$ is in a different color representation as compared to the initial one of the line at $\xvec_1$ in the test operator. Namely irreps $R_{\ell/r}$ are given as a function of $R_1$ by
\begin{align}
R_\ell \equiv \bar{F} \, \delta_{R_1 = A} + A \, \delta_{R_1 = F} \\ 
R_r \equiv F \, \delta_{R_1 = A} + A \, \delta_{R_1 = \bar{F}}
\end{align}
and the system crossing the shock wave is $R_0 \otimes F \otimes R_\ell$ or $R_0 \otimes \bar{F} \otimes R_\ell$.

We saw earlier how the action of the operator \eqref{EQ:Diff_op_psi_bpsi} on a polarized Wilson line in the irrep $R_1$ generates a Wilson line crossing the shock-wave in a representation different from $R_1$. In Eqs.~\eqref{EQ:V_pol_d_bpsi_psi} and \eqref{EQ:U_pol_d_bpsi_psi}, after setting $y_1^-$ and $y_2^-$ to be on different sides of the shock-wave, the Wilson line that crossed the shock-wave was in the representations $R_\ell$ or $R_r$, both distinct from $R_1$.
As expected from \fig{FIG:Evolution_psi_field}, we first act on our test operator with the double functional derivative \eqref{EQ:Diff_op_psi_bpsi}, then project out the helicity-dependent part using \eq{hel_proj}, and, finally, contract the result of those operations with the polarized Wilson line at the position $\wvec$ to recover the evolution kernel. 
Fixing the overall coefficient by using the explicit calculations in \cite{Kovchegov:2018znm} we arrive at 
\begin{align}
\calk^{\psi,\bar{\psi}} = (x\cdot y) &\left\{  \left(\tfrac{1}{2}\gamma^5\gamma^-\right)_{\beta\alpha} \tfrac{1}{2} \left( (V_\wvec^{pol})_{ij}  D^{\bar{\psi}}_{j, \alpha,<} D^{{\psi}}_{i, \beta,>} \right.\right. \ \nonumber \\
&\qquad \quad \left. \left. +\ (V_\wvec^{pol^\dagger})_{ij}  D^{\bar{\psi}}_{j, \alpha,>} D^{{\psi}}_{i, \beta,<} \right) \right\},
\end{align}
where we again suppressed the position subscripts on the functional derivatives for brevity.

Those functional derivatives are understood to acts only on the polarized line at $\xvec_1$. Just like for $\calk^\beta$, one can make the kernel $\calk^{\psi,\bar{\psi}}$ symmetric under the $x \leftrightarrow y$ interchange by allowing those derivatives to act on both Wilson lines in \eq{EQ:Test_operator_pol}  since the unpolarized Wilson line at $\xvec_0$ does not depend on the two quark fields $\{ \psi, \bar{\psi}\}$.

\smallbreak
\paragraph{The result for the kernel.}
The full evolution kernel for the test operator $\hcalo_{\ov,\zv}$ reads 
\begin{equation}
\label{Eq:Small_heliity_kernel}
\calk_h = \calk^{eik} + \calk^\beta + \calk^{\psi,\bar{\psi}},
\end{equation}
that is,
\begin{widetext}
\begin{align}
\label{Eq:Full_helicity_Kernel}
& \calk_h [\tau, \tau']  \equiv \frac{\alpha_s}{\pi^2} \int \dd w_\perp
\frac{{\un X}' \cdot {\un Y}'}{X'^2 \, Y'^2}\ 
\theta^{(3)}(\tau-\tau') \theta \left( z' - \tfrac{\Lambda^2}{s} \right) \, \theta \left( X'^2 - \tfrac{1}{z' \, s} \right) \, \theta \left( Y'^2 - \tfrac{1}{z' \, s} \right) \\ & \times \, \left\{
U_\wvec^{ba} D^+_{\xvec, a, <} \, D^+_{\yvec, b, >} - \tfrac{1}{2} \left( D^+_{\xvec, a, <} D^+_{\yvec, a, <} + D^+_{\xvec, a, >} \, D^+_{\yvec, a,>} \right) \right. \notag  \\
& \left. + \tfrac{1}{2}  U^{pol,ba}_\wvec ( D^+_{\xvec, a, <} D^\perp_{\yvec, b, >} + D^\perp_{\xvec, a, <} D^+_{\yvec, b, >} ) + \left(\tfrac{1}{2}\gamma^5\gamma^-\right)_{\beta\alpha} \tfrac{1}{2} \left((V_\wvec^{pol})_{ij}  D^{\bar{\psi}}_{\xvec, j, \alpha,<} D^{{\psi}}_{\yvec, i, \beta,>} + (V_\wvec^{pol \, \dagger})_{ij}  D^{\bar{\psi}}_{\xvec, j, \alpha,>} D^{{\psi}}_{\yvec, i, \beta,<} \right) 
\right\}. \nonumber
\end{align}
\end{widetext}
It should be noted that the number of quark flavors was suppressed here (i.e., $N_f=1$). To bring it back into \req{Eq:Full_helicity_Kernel} one introduces flavor dependence for the polarized quark Wilson lines, 
\begin{align}
V_{\un x}^{pol} \to V_{\un x}^{pol, \, f} ,
\end{align}
where the only modification compared to \eq{EQ:V_pol} is that the quark-exchange contribution to $V_{\un x}^{pol, \, f}$ is given by \eq{EQ:V_pol_q} with flavor-specific quark fields, $\{ \psi, {\bar \psi} \} \to \{ \psi^f , {\bar \psi}^f \}$. The adjoint polarized Wilson line $U_{\un x}^{pol}$ is still given by \eq{EQ:U_pol} with the following modification: the quark-exchange contribution to the adjoint polarized Wilson line should be given by \eq{EQ:U_pol_q} with a sum over exchanged quark flavors,
\begin{align}
{\bar \psi} \ldots \psi \to \sum_f {\bar \psi}^f \ldots \psi^f .  
\end{align}
Naturally the adjoint polarized Wilson lines do not acquire a flavor index. The kernel $\calk_h$ is modified by summing the last term in the curly brackets over flavors and introducing the quark flavor index in the functional derivatives over fermions fields,
\begin{subequations}
\begin{align}
& (V_\wvec^{pol})_{ij}  D^{\bar{\psi}}_{\xvec, j, \alpha,<} D^{{\psi}}_{\yvec, i, \beta,>} \!\! \to \sum_f (V_\wvec^{pol, \, f})_{ij}  D^{\bar{\psi}^f}_{\xvec, j, \alpha,<} D^{{\psi}^f}_{\yvec, i, \beta,>},  \\ 
& (V_\wvec^{pol \, \dagger})_{ij}  D^{\bar{\psi}}_{\xvec, j, \alpha,>} D^{{\psi}}_{\yvec, i, \beta,<} \to \sum_f  (V_\wvec^{pol\, f \, \dagger})_{ij}  \\ & \hspace*{5cm} \times D^{\bar{\psi}^f}_{\xvec, j, \alpha,>} D^{{\psi}^f}_{\yvec, i, \beta,<}. \notag
\end{align} 
\end{subequations}
This accomplishes the generalization of our formalism to an arbitrary number of quark flavors. 

In addition, as we have already pointed out above for the eikonal contribution to the kernel, the transverse integrals in the expression one obtains by inserting \eq{Eq:Full_helicity_Kernel} into \eq{EQ:Operator_evo_helicity1} need to be simplified further to extract the logarithmic contribution giving the second $\ln (1/x)$ in the DLA, as it was done in \cite{Kovchegov:2015pbl,Kovchegov:2017lsr,Kovchegov:2018znm}.  

Finally, let us note that, as was the case with the JIMWLK kernel, this \textit{flavor-singlet helicity kernel} applies to evolution of a broad range of test flavor-singlet operators consisting of any number of Wilson lines, with one polarized Wilson line, in any irrep $R$. This is in addition to this kernel containing the eikonal part, which generates the standard JIMWLK or BK evolution for the operators made out of regular light-cone Wilson lines. \\


\subsection{Cross-check with BER in the ladder approximation}\label{Sec:3b}

In this section we cross-check our results with Eq.~(64-66) of \cite{Kovchegov:2015pbl} to verify whether, in the ladder approximation, the kernel given by \req{Eq:Full_helicity_Kernel} coincides with that obtained by Bartels, Ermolaev and Ryskin (BER) \cite{Bartels:1996wc,Bartels:1995iu}.

First, we consider a test operator consisting of a doublet of color-singlet polarized dipoles in representations $F$ and $A$. Let us fix the polarized Wilson line at the position $\xvec_1$. After that, we set all unpolarized Wilson lines to be identities in the color space: we can do this if we are only interested in the ladder contribution \cite{Kovchegov:2015pbl}. 
As a result, the test operator reduces to \cite{Kovchegov:2015pbl}
\begin{equation}\label{W_op}
\hcalo_{pol} \rightarrow 
\begin{pmatrix}
\frac{1}{K_F} \left[ \tr(V^{pol}_\ov) + \tr(V^{pol,\dagger}_\ov) \right] \\
\frac{1}{K_A} \left[ \Tr(U^{pol}_\ov) + \Tr(U^{pol,\dagger}_\ov) \right]
\end{pmatrix}
= W_{\ov,\zv}^{pol} .
\end{equation}
The normalization is chosen as in Eq.~(64) of \cite{Kovchegov:2015pbl}: irrep's dimensions are $K_F = N_c$ and $K_A = N_c^2 -1$. We can obtain the DLA evolution of the operator \eqref{W_op} by using the kernel $\calk_h$ from \eq{Eq:Full_helicity_Kernel} in \eq{EQ:Operator_evo_helicity1}. To truncate this evolution to the level containing ladder diagrams only, we will only keep emissions generating softer polarized lines. That is, we will discard $\calk^{eik}$ from $\calk_h$ (see \eqref{Eq:Small_heliity_kernel}), and keep only $\calk^\beta$ and $\calk^{\psi,\bar{\psi}}$. Indeed there is no formal justification for this ladder approximation: as stressed in \cite{Kovchegov:2015pbl}, it is simply a truncation needed to compare a part of our calculation with the same part of BER. 

In order to recast \eq{EQ:Operator_evo_helicity1} in a form closer to Eq.~(65) in \cite{Kovchegov:2015pbl}, we change the integration variable from $\wvec$ to $\wvec - \xvec_1$, integrate over the angles of $\wvec - \xvec_1$, and keep the integral over $|\wvec - \xvec_1|^2$.

Anticipating all derivatives to act on the polarized Wilson lines at $\un 1$, we rewrite \eq{EQ:Operator_evo_helicity1} for the modified test operator \eqref{W_op} as 
\begin{widetext}
\begin{align}\label{KhO}
& \left\langle W_{\ov,\zv}^{pol} \right\rangle_\tau =  \left\langle W_{\ov,\zv}^{pol} \right\rangle_0 + \frac{\alpha_s}{2\pi} \int\limits_{\Lambda^2/s}^{z} \frac{\text{d}z'}{z'}\int\limits^{\xvec_{10}^2 z/z'}_{1/z's} \frac{\text{d} |\wvec - \xvec_1|^2}{|\wvec - \xvec_1|^2}  \int d^2 x_\perp \, d^2 y_\perp
 \\
& \times \, \left\langle 2 \, \left\{
\tfrac{1}{2}  U^{pol,ba}_\wvec ( D^+_{\xvec, a, <} D^\perp_{\yvec, b, >} + D^\perp_{\xvec, a, <} D^+_{\yvec, b, >} ) + \left(\tfrac{1}{2}\gamma^5\gamma^-\right)_{\beta\alpha} \tfrac{1}{2} \left((V_\wvec^{pol})_{ij}  D^{\bar{\psi}}_{\xvec, j, \alpha,<} D^{{\psi}}_{\yvec, i, \beta,>} + \mbox{h.c.}  \right) 
\right\} W_{\ov,\zv}^{pol} \right\rangle .  \nonumber
\end{align}
\end{widetext}
The matrix $M$ from Eq.~(66) in \cite{Kovchegov:2015pbl} is obtained by estimating $2 \{ \cdots \}$ in \eq{KhO} acting on $W_{\ov,\zv}^{pol}$.
The factor we obtain when acting with the first term in the curly brackets on the right of \eq{KhO} onto a polarized Wilson line in an arbitrary representation $R$ is
\begin{equation}\label{check1}
2\frac{\eta_R a_R K_A}{K_R} = 2C_F \, \delta_{R=F} + 4N_c \, \delta_{R=A},
\end{equation}
where $a_R$ is the normalization of the trace of two color generator in irrep R, i.e., $\tr [t_R^a \, t_R^b] = a_R \, \delta^{ab}$, such that $C_R K_R = a_R K_A$, and the factor $\eta_R$ is defined in \req{eta_def} as $\eta_R = \delta_{R=F} + \delta_{R=\bar{F}} + 2 \delta_{R=A}$.
The second term in the curly brackets of \eq{KhO} gives
\begin{equation}\label{check2}
\frac{2C_F K_F}{K_R} = C_F\, \delta_{R=F}  -  \delta_{R=A} \rightarrow C_F\, \delta_{R=F}  - N_f \delta_{R=A},
\end{equation}
where we have restored the number-of-flavors factor $N_f$ on the right-hand side of \eq{KhO}.

The factors multiplying the Kronecker deltas in Eqs.~\eqref{check1} and \eqref{check2} give the coefficients in the matrix $M$ (or the matrix $M_0$ from Eq~(2.28) in \cite{Bartels:1996wc}), thus completing the cross-check with the ladder limit of BER.


\section{Weight functional evolution equation}
\label{Sec:4}

Having obtained the evolution kernel for the test operator, one just needs to integrate \req{EQ:Operator_evo_helicity1}  by parts with $\calk_h$ given by \req{Eq:Full_helicity_Kernel} to obtain the helicity-dependent version of the JIMWLK evolution equation for the target weight functional. That is, we now need to show that \eq{EQ:Evolution} really follows from \req{EQ:Operator_evo_helicity1}.


\subsection{Acting on the weight functional with helicity evolution kernel} 

The procedure will be the same for each term of the evolution kernel $\calk_h$: (i) fix the irrelevant fields and consider the average over configurations of the relevant fields, (ii) perform the integration by parts over the relevant fields, and (iii) state the required constraints on the integration boundaries.

For brevity, in the following only the functional integrals over relevant field are shown explicitly. We will carry out the integration by parts term-by-term in \eq{Eq:Small_heliity_kernel}.

\smallbreak
\paragraph{Eikonal kernel.}
The integration by parts of the eikonal kernel $\calk^{eik}$  is completely analogous to the one in the JIMWLK equation. For a set of configurations with fixed sub-eikonal fields $\beta$, $\psi$, and $\bpsi$, it gives
\begin{equation}
\int \cald \alpha \  d^3 \tau' \  \hcalo_{pol} \ \left( \calk^{eik}[\tau, \tau']  \cdot \mathcal{W} \right)[\alpha] .
\end{equation}
Here we require the surface term to vanish on the boundary, that is at $\alpha (x) \rightarrow \pm \infty$, thus obtaining a constraint that the weight functional should vanish for infinite functions $\alpha (x)$.


\smallbreak
\paragraph{Sub-eikonal kernel: the fermion fields.}
Let us consider the $\calk^{\psi,\bar{\psi}}$ part of the kernel $\calk_h$. For a set of configurations with fixed $\alpha$ and $\beta$, we want to integrate
\begin{align}
\int \cald \psi\, \cald\bar{\psi} \ d^3 \tau' \  & (x\cdot y) \left[ \thalf \left( V^{pol}_{\un w}  D_<^\bpsi D_>^\psi + h.c. \right)\, \hcalo_{pol} \right] \notag \\ & \times \, \mathcal{W}[\psi,\bpsi] , 
\end{align}
by parts, where, for brevity, we have defined
\begin{align}
V^{pol}_{\un w}  D_<^\bpsi D_>^\psi \equiv \left(\tfrac{1}{2}\gamma^5\gamma^-\right)_{\beta\alpha}  (V_\wvec^{pol})_{ij}  D^{\bar{\psi}}_{\xvec, j, \alpha,<} D^{{\psi}}_{\yvec, i, \beta,>} .
\end{align} 
Using the fact that for any two complex-valued functions $f (\theta, \bar{\theta})$ and $g (\theta, \bar{\theta})$ of a complex Grassmann number $\theta$ (with $\bar{\theta}$ denoting its complex conjugate) one has
\begin{equation}
\label{EQ:IbPGrassmann}
\int d\theta d\bar{\theta} \ f \ \frac{d^2 g}{d\bar{\theta} d\theta} = \int d\theta d\bar{\theta}\ \frac{d^2 f}{d\bar{\theta} d\theta}\ g
\end{equation}
and the fact that the functional derivative acting on $V^{pol}_{\un w}$ vanishes,
\begin{equation}
D_{j, <}^{\bar{\psi}} \left(V^{pol}_{\un w} \right)_{ij} = 
D_{i, >}^{\psi} 
\left(V^{pol}_{\un w} \right)_{ij} = 0 
\end{equation}
(if we put the unpolarized Wilson lines outside the shock wave to unity), one easily obtains
\begin{equation}
\int \cald \psi\, \cald \bpsi \ d^3 \tau' \  \hcalo_{pol}  \left(\calk^{\psi,\bpsi}[\tau, \tau']  \cdot  \mathcal{W}\right)[\psi,\bpsi]. 
\end{equation}
There is no constraint on the boundaries here since there is no surface term in \req{EQ:IbPGrassmann}.


\smallbreak
\paragraph{Sub-eikonal kernel: the $\beta$-field.}
The remaining part of the kernel $\calk_h$ in \eqref{Eq:Small_heliity_kernel} that we need to integrate by parts is $\calk^\beta$.
Fixing the irrelevant fields $\psi$ and $\bpsi$, we consider the average
\begin{align}
\int \cald \alpha \cald\beta \ d^3 \tau' \ & (x\cdot y)  \left[\thalf U_\tv^{pol} \left( D_{<}^+ D_{>}^\perp + D_{<}^\perp D_{>}^+ \right) \hcalo_{pol} \right] \notag \\ & \times \, \mathcal{W}[\alpha,\beta] .
\end{align}
Integrating by parts and using (again, after putting the unpolarized Wilson lines outside the shock wave to unity)
\begin{equation}
D_{a,<}^{+,\perp} (U^{pol})^{ba} = D_{b,>}^{+,\perp} (U^{pol})^{ba} = 0
\end{equation}
one can write
\begin{equation}
\int \cald \alpha \cald\beta \ d^3 \tau' \  \hcalo_{pol} \, \left(\calk^\beta [\tau, \tau'] \cdot \mathcal{W}\right)[\alpha,\beta] .
\end{equation}
We require that the surface terms vanish on the boundary, that is, at  $\alpha (x) \rightarrow \pm \infty$ and $\beta(x) \rightarrow \pm \infty$. This gives an additional constraint on the weight functional: it has to vanish at $\beta(x) \rightarrow \pm \infty$.


\smallbreak
\paragraph{The result.}
The above procedures finally yield the flavor-singlet helicity-dependent generalization of the JIMWLK evolution equation for the target weight functional (cf. \eq{EQ:Evolution}),
\begin{align}
\label{EQ:Helicity_evolution_equation}
{\cal W}_\tau [\alpha,\beta,\psi,\bpsi] = &\ \calw_\tau^{(0)} [\alpha,\beta,\psi,\bpsi]  \\ 
& + \int d^3 \tau' \ \calk_h[\tau,\tau'] \cdot {\cal W}_{\tau'}  [\alpha,\beta,\psi,\bpsi] \nonumber
\end{align}
with the kernel $\calk_h$ given by \req{Eq:Full_helicity_Kernel}. Equations \eqref{EQ:Helicity_evolution_equation} and \eqref{Eq:Full_helicity_Kernel} are the main result of this work.


\subsection{Properties of the solution and the initial condition}

Here we infer some properties of the weight functional $\calw_\tau$ given by the solution of the helicity evolution equation \eqref{EQ:Helicity_evolution_equation} without solving the latter explicitly. We also determine the properties we expect the inhomogeneous term $\calw_\tau^{(0)}$ to have. 

Due to the sub-eikonal origin of the flavor-singlet helicity evolution equation, we expect the weight functional $\mathcal{W}_\tau$ to respect the following ansatz:
\begin{equation}\label{ansatz}
\calw_\tau = \mathcal{W}_\tau^{unpol} + \Sigma \, \mathcal{W}_\tau^{pol},
\end{equation}
where the labels \textit{pol} or \textit{unpol} indicate whether the functional enters $\calw_\tau$ with or without a factor of the target helicity $\Sigma$. 
This ansatz is similar to how helicity evolution enters other observables \cite{Kovchegov:2015pbl,Kovchegov:2016zex,Kovchegov:2016weo,Kovchegov:2017jxc,Kovchegov:2017lsr,Kovchegov:2018znm}.  We expect the decomposition \eqref{ansatz} to be valid for the inhomogeneous term too,
\begin{equation}\label{ansatz0}
\calw^{(0)}_\tau = \mathcal{W}^{(0) \, unpol} + \Sigma \, \mathcal{W}_\tau^{(0) \, pol}.
\end{equation}
The inhomogeneous terms are calculated from the initial conditions for the evolution.

Further, requiring that the ansatz \eqref{ansatz} maps back onto the standard JIMWLK evolution after the sub-eikonal fields have been integrated out yields 
\begin{subequations}
\begin{align}
\hspace{-3mm}   \int \cald\beta \cald\psi \cald \bar{\psi} \ \mathcal{W}^{unpol}_\tau [\alpha,\beta,\psi,\bpsi] &\,  = \mathcal{W}_\tau^{JIMWLK}[\alpha],  \label{W_JIMWLK} \\
\int \cald\beta \cald\psi \cald \bar{\psi} \ \mathcal{W}^{pol}_\tau [\alpha,\beta,\psi,\bpsi] &\,  = 0 , \label{Wpol_sum}
\end{align}
\end{subequations}
along with
\begin{subequations}
\begin{align}
\int \cald\beta \cald\psi \cald \bar{\psi} \ \mathcal{W}^{(0) \, unpol} [\alpha,\beta,\psi,\bpsi] &\,  = \mathcal{W}^{MV}[\alpha],  \label{W_MV} \\
\int \cald\beta \cald\psi \cald \bar{\psi} \ \mathcal{W}^{(0) \, pol}_\tau [\alpha,\beta,\psi,\bpsi] &\, = 0 , \label{Wpol_0}
\end{align}
\end{subequations}
for the inhomogeneous term. 
Here $\mathcal{W}^{JIMWLK}[\alpha]$ is the standard JIMWLK-evolved weight functional \cite{Jalilian-Marian:1997dw,Jalilian-Marian:1997gr,Weigert:2000gi,Iancu:2001ad,Iancu:2000hn,Ferreiro:2001qy} while $\mathcal{W}^{MV}[\alpha]$ is the Gaussian weight functional of the McLerran--Venugopalan (MV) model \cite{McLerran:1993ni,McLerran:1993ka,McLerran:1994vd,Kovchegov:1996ty}. 

Substituting the ansatze \eqref{ansatz} and \eqref{ansatz0} into \eq{EQ:Helicity_evolution_equation} and separating the target helicity $\Sigma$ dependent and independent terms, we arrive at two equations,
\begin{subequations} 
\begin{align}
{\cal W}_\tau^{JIMWLK} [\alpha] = &\ \calw^{MV} [\alpha]  \label{JIMWLK_t} \\ 
& + \int d^3 \tau' \ \calk^{eik} [\tau,\tau'] \cdot {\cal W}^{JIMWLK}_{\tau'}  [\alpha] , \nonumber \\
{\cal W}^{pol}_\tau [\alpha,\beta,\psi,\bpsi] = &\ \calw_\tau^{(0) \, pol } [\alpha,\beta,\psi,\bpsi]   \label{hJIMWLK} \\ 
& + \int d^3 \tau' \ \calk_h[\tau,\tau'] \cdot {\cal W}^{pol}_{\tau'}  [\alpha,\beta,\psi,\bpsi],  \notag
\end{align}
\end{subequations} 
where the first equation we have also integrated over $\beta, \psi$ and $\bar \psi$ using Eqs.~\eqref{W_JIMWLK} and \eqref{W_MV} . Equation \eqref{JIMWLK_t} is the usual JIMWLK equation, written in the integral form. Note that for this unpolarized evolution, all transverse integrals are convergent both in the infrared and in the ultraviolet: hence, the lifetime ordering conditions present in the kernel of \eq{Keik1} are not important at the leading-logarithmic in $\ln (1/x)$ order (LLA), and do not affect the identification of \eq{JIMWLK_t} as the standard JIMWLK equation. It is curious, though, that these lifetime ordering conditions need to be present in the (generalized) leading-order JIMWLK kernel already to give us correct helicity evolution: it has been argued recently that such lifetime orderings may be important at the next-to-leading logarithmic (NLL) order in the unpolarized small-$x$ BK evolution \cite{Iancu:2015vea,Ducloue:2019ezk}. Using our result \eqref{Eq:Full_helicity_Kernel} for $\calk_h$ one could speculate that lifetime ordering needs to be imposed in all shock wave evolution calculations, including the standard JIMWLK and BK evolution, and not just for the DLA helicity evolution we consider here. This ordering may help one better organize calculations of the higher-order corrections to the small-$x$ evolution kernel. 

Equation \eqref{hJIMWLK} is the functional equation for helicity evolution. Note that if one assumes that the inhomogeneous term for this equation satisfies the condition \eqref{Wpol_0}, then, integrating \eq{hJIMWLK} over $\beta, \psi$ and $\bar \psi$ one can show that \eq{Wpol_sum} is satisfied by our evolution. 

Next imagine using the weight functional $\calw_\tau$ from \eq{ansatz} to calculate an expectation value of some polarization-dependent operator, say, the scattering amplitude of a high-energy quark in the target field with the eikonal and the sub-eikonal helicity-dependent terms included \cite{Kovchegov:2015pbl,Kovchegov:2016zex,Kovchegov:2017lsr,Kovchegov:2018znm}, 
\begin{align}
V_\xvec (\sigma) = V_\xvec + \sigma \, V_\xvec^{pol}. 
\end{align}
Its expectation value is
\begin{align}\label{aveV1}
& \left\langle V_\xvec (\sigma) \right\rangle_\tau = \int \cald\alpha \cald\beta \cald\psi \cald \bar{\psi} \ {\cal W}_\tau [\alpha,\beta,\psi,\bpsi] \, V_\xvec (\sigma) \\ & = \int \cald\alpha \cald\beta \cald\psi \cald \bar{\psi} \ \left[ \mathcal{W}_\tau^{unpol} + \Sigma \, \mathcal{W}_\tau^{pol} \right] \, \left[ V_\xvec + \sigma \, V_\xvec^{pol} \right]. \notag
\end{align}
Due to PT symmetry, only the term independent of helicities and the term proportional to $\sigma \, \Sigma$ should survive, 
\begin{align}\label{aveV2}
\left\langle V_\xvec (\sigma) \right\rangle_\tau = & \, \int \cald\alpha \cald\beta \cald\psi \cald \bar{\psi} \\ \times & \  \left[ \mathcal{W}_\tau^{unpol} \, V_\xvec + \sigma \, \Sigma \, \mathcal{W}_\tau^{pol} \, V_\xvec^{pol} \right]. \notag
\end{align}
For the terms containing either only $\sigma$ or only $\Sigma$ in \eq{aveV1} to vanish, we need 
\begin{subequations} 
\begin{align}
& \int \cald\alpha \cald\beta \cald\psi \cald \bar{\psi} \ \mathcal{W}_\tau^{unpol} [\alpha,\beta,\psi,\bpsi] \, V_\xvec^{pol} (\alpha,\beta,\psi,\bpsi)  = 0 , \label{WVpol} \\
& \int \cald\alpha \cald\beta \cald\psi \cald \bar{\psi} \ \mathcal{W}_\tau^{pol} [\alpha,\beta,\psi,\bpsi]  \, V_\xvec (\alpha) =0. \label{WpolV}
\end{align}
\end{subequations} 
While the condition \eqref{WpolV} easily follows from \eq{Wpol_sum}, \eq{WVpol} appears to yield a new condition, likely constraining the dependence of $\mathcal{W}_\tau^{unpol} $ on the sub-eikonal fields $\beta, \psi$ and $\bar \psi$.

In our derivation above we have employed the fact that the weight functional $\calw_\tau$ should vanish on the boundary $\alpha (x), \beta (x) \rightarrow \pm \infty$, that is,
\begin{equation}
\begin{cases}
\lim\limits_{\alpha(x) \rightarrow \pm \infty} \calw_\tau [\alpha,\beta,\psi,\bpsi] = 0, \qquad \forall x , \\ 
\lim\limits_{\beta(x) \rightarrow \pm \infty} \calw_\tau [\alpha,\beta,\psi,\bpsi] = 0, \qquad \forall x .
\end{cases}
\end{equation} 
The initial weight functional $\calw_\tau^{(0)}$ should satisfy the same boundary conditions, 
\begin{equation}\label{W0boundary}
\begin{cases}
\lim\limits_{\alpha(x) \rightarrow \pm \infty} \calw_\tau^{(0)}[\alpha,\beta,\psi,\bpsi] = 0, \qquad \forall x , \\ 
\lim\limits_{\beta(x) \rightarrow \pm \infty} \calw_\tau^{(0)}[\alpha,\beta,\psi,\bpsi] = 0, \qquad \forall x .
\end{cases}
\end{equation} 
The former boundary condition is expected from the MV model \cite{McLerran:1993ni,McLerran:1993ka,McLerran:1994vd}, where the weight functional has a Gaussian form \cite{Kovchegov:1996ty}. This Gaussian form is not obvious {\sl a priori} for the helicity-dependent term, and motivating this behavior is the topic of our next paper \cite{CK2}. Verification of the ansatz \eqref{ansatz0}, along with construction of an explicit expression for $\calw_\tau^{(0)}$ and cross-checking that it satisfies the boundary conditions \eqref{W0boundary} are left for the future work \cite{CK2}.


\section{Conclusions and Outlook}
\label{Sec:5}

In this paper we have derived a helicity-dependent extended version of the JIMWLK evolution equation. The result is given by Eqs.~\eqref{EQ:Helicity_evolution_equation} and \eqref{Eq:Full_helicity_Kernel}, and applies for DLA helicity evolution of the flavor-singlet operators in addition to the LLA evolution of the standard operators made out of light-cone Wilson lines. 

As mentioned above, this equation has several practical applications: (i) it can be used to write down evolution equations for any sub-eikonal helicity-dependent flavor-singlet operator, made out of any number of standard Wilson lines and one polarized Wilson line; (ii) it may be possible to solve equation \eqref{EQ:Helicity_evolution_equation} numerically, allowing one to determine the small-$x$ asymptotics of any operator of the type described in (i) beyond the large-$N_c$ and large-$N_c \,\&\, N_f$ limits considered in \cite{Kovchegov:2015pbl,Kovchegov:2016zex,Kovchegov:2017lsr,Kovchegov:2018znm}. Both applications are important for identifying the small-$x$ asymptotics of quark and gluon helicity distributions, to be measured in the future experiments such as at the proposed Electron-Ion Collider (EIC) \cite{Accardi:2012qut}. 

To complete \eq{EQ:Helicity_evolution_equation} one needs to determine the inhomogeneous term $\calw_\tau^{(0)}$ in that equation. This is left for future work \cite{CK2}. The exact form of this inhomogeneous term is not likely to affect the small-$x$ asymptotics of helicity distributions (see \cite{Kovchegov:2016weo} where it was demonstrated that the small-$x$ helicity evolution in the large-$N_c$ limit is not very sensitive to its initial conditions given by the inhomogeneous term). Nevertheless, the inhomogeneous term may be important for the comparison of helicity evolution to data at small but finite values of $x$.  

In the future, the JIMWLK approach can also be applied to the case of helicity evolution for the flavor non-singlet operators. It appears that to achieve this one needs to go beyond the polarized quark and gluon Wilson line operators in Eqs.~\eqref{EQ:V_pol} and \eqref{EQ:U_pol}: in addition, one needs to introduce operators where an incoming $s$-channel quark transitions into an $s$-channel gluon after a non-eikonal helicity-dependent interaction (along with the reverse process where an $s$-channel gluon transitions into an $s$-channel quark). Diagrammatically these operators correspond to the right two graphs in \fig{FIG:SubEik_int_vertices}, further dressed by multiple eikonal gluon exchanges with the target \cite{Kovchegov:2017lsr,Kovchegov:2018znm}. The resulting flavor non-singlet helicity analogue of JIMWLK evolution would allow one to determine the small-$x$ asymptotics of the flavor non-singlet helicity distributions beyond the large-$N_c$ limit studied in \cite{Kovchegov:2016zex}. Similar to the flavor-singlet case, this may have possible phenomenological applications. It would also allow one to compare a prediction of the shock-wave formalism for this flavor non-singlet observable to the results based on other approaches \cite{Bartels:1995iu} beyond the large-$N_c$ limit.  

Before we conclude, let us reiterate another very intriguing feature of our result. The kernel \eqref{Eq:Full_helicity_Kernel} generates both the standard unpolarized LLA small-$x$ evolution and DLA helicity evolution. The eikonal part of this kernel is the usual LLA JIMWLK kernel, which, to get the correct DLA helicity evolution, had to be augmented by the $\theta$-functions imposing the lifetime ordering in the $x^-$ light-cone direction. These $\theta$-functions do not affect the LLA unpolarized evolution, but may become important for the higher-order corrections to the BK evolution kernel, as was argued in \cite{Iancu:2015vea,Ducloue:2019ezk}. Their presence reduces the magnitude of the NLL and higher-order corrections by generating a ``rapidity veto" in the BK evolution. It is quite curious that such lifetime ordering corrections appear in the eikonal part of our kernel \eqref{Eq:Full_helicity_Kernel} due to a formal requirement that it generates correct DLA helicity evolution: this appears to be an independent formal argument in favor of keeping lifetime ordering to all orders of perturbation theory in the JIMWLK and BK kernels.


\section*{Acknowledgments}

One of the authors (YK) would like to thank Raju Venugopalan for a discussion of the concept of a JIMWLK analogue for helicity evolution. YK is also indebted to Ian Balitsky for several discussions of the inverse-ordering emissions and their cancellation by the shock wave corrections, as discussed in Appendix~\ref{app}. This material is based upon work supported by the U.S. Department of
Energy, Office of Science, Office of Nuclear Physics under Award
Number DE-SC0004286.



\appendix
\section{On the simultaneous light-cone momentum fraction ordering and lifetime ordering in unpolarized and polarized small $x$ evolution}
\label{app}

\begin{figure}[ht]
\begin{center}
\includegraphics[width= 0.3 \textwidth]{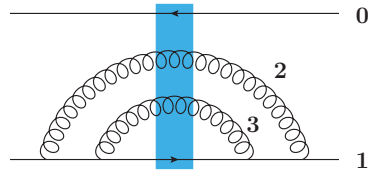} 
\caption{Emission of two gluons in a dipole as considered in the text.}
\label{gluon23}
\end{center}
\end{figure}

Here we would like to show that in the DLA helicity evolution the emissions have to be ordered both in the light-cone momentum fraction $z$ and in the parton lifetime $z \, x_\perp^2$. Specifically, consider emitting gluons $2$ and $3$ in a quark--antiquark dipole $10$, as shown by one possible contributing diagram in \fig{gluon23}. At this point our discussion is quite general and the emissions can be either polarized or unpolarized. 

It is easy to see that in order to obtain the leading-logarithmic (LLA) contribution from the gluons pictured in \fig{gluon23} for the case of unpolarized evolution, or the leading DLA contribution for helicity evolution, the energy denominators of the two gluons in \fig{gluon23} have to be ordered. (We are using the terminology of the light-cone perturbation theory (LCPT) \cite{Lepage:1980fj}.) This condition leads to the lifetime ordering, 
\begin{align}\label{Eordering}
z_2 \, x_{21}^2 \gg z_3 \, x_{31}^2, 
\end{align}
where, as usual, $x_{ij} = |{\un x}_i - {\un x}_j|$.

\begin{figure*}
\begin{center}
\includegraphics[width= 0.9 \textwidth]{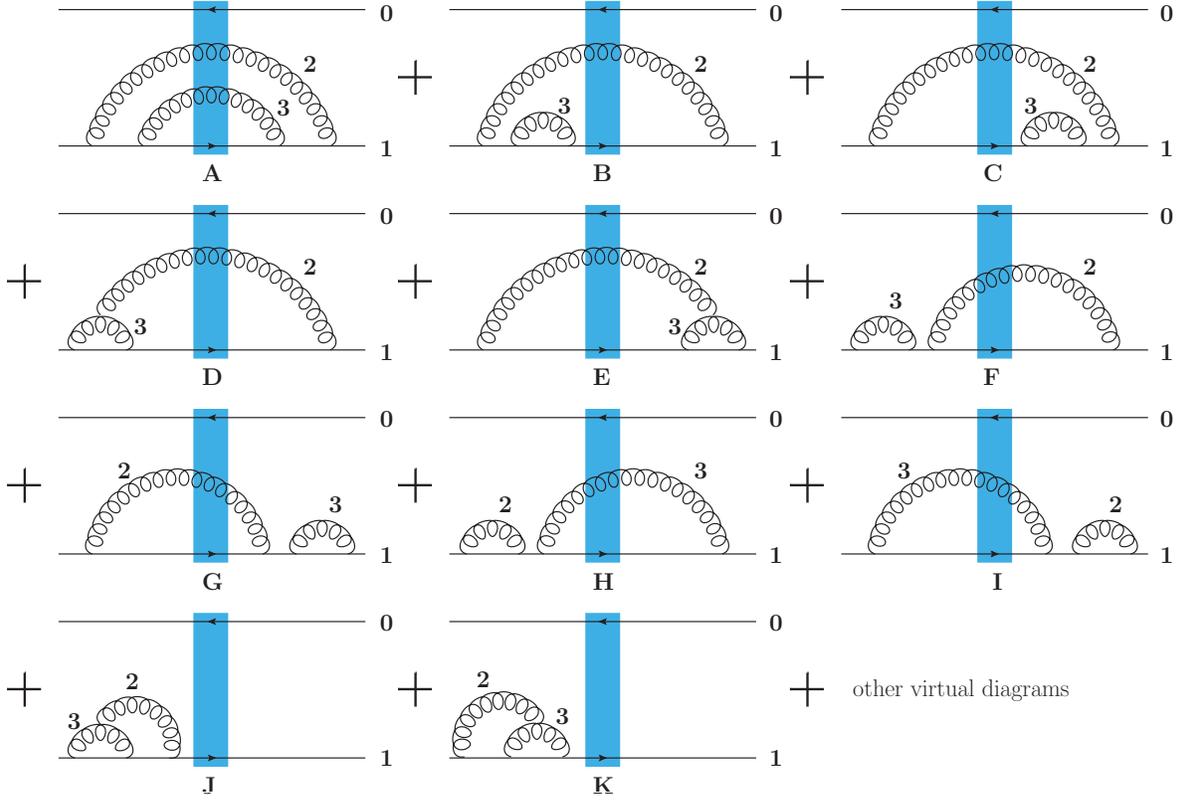} 
\caption{The large-$N_c$ diagrams contributing to emission of two gluons (2,3) in a dipole 10 in the kinematics of \eq{Ezordering} for the unpolarized evolution.}
\label{unpol}
\end{center}
\end{figure*}

In the case of unpolarized evolution \cite{Balitsky:1995ub,Balitsky:1998ya,Kovchegov:1999yj,Kovchegov:1999ua,Jalilian-Marian:1997dw,Jalilian-Marian:1997gr,Weigert:2000gi,Iancu:2001ad,Iancu:2000hn,Ferreiro:2001qy} and for helicity evolution of \cite{Kovchegov:2015pbl,Kovchegov:2016zex,Kovchegov:2017lsr,Kovchegov:2018znm}, one usually imposes the light-cone momentum fraction ordering such that 
\begin{align}\label{zordering}
z_2 \gg z_3. 
\end{align}
For the unpolarized evolution, where one assumes that all transverse distances are comparable, $x_{21} \sim x_{31}$, this second condition \eqref{zordering} appears to follow from \eq{Eordering}. For the helicity evolution the two conditions \eqref{Eordering} and \eqref{zordering} are separate since in helicity evolution the transverse distance integrals lead to logarithms of energy, coming, in part, from very small dipole sizes: hence the $x_{21} \sim x_{31}$ assumption may not always be valid. 

It is, therefore, important to explore the case of the ``opposite" or ``inverse" ordering of the light-cone momentum fractions, $z_2 \ll z_3$, imposed simultaneously with the ``correct" lifetime ordering \eqref{Eordering}. That is, we should consider the case of
\begin{align}\label{Ezordering}
z_2 \, x_{21}^2 \gg z_3 \, x_{31}^2, \ \ \ z_2 \ll z_3 .
\end{align}
Both the unpolarized \cite{Balitsky:1995ub,Balitsky:1998ya,Kovchegov:1999yj,Kovchegov:1999ua,Jalilian-Marian:1997dw,Jalilian-Marian:1997gr,Weigert:2000gi,Iancu:2001ad,Iancu:2000hn,Ferreiro:2001qy} and helicity \cite{Kovchegov:2015pbl,Kovchegov:2016zex,Kovchegov:2017lsr,Kovchegov:2018znm} evolutions assume the ``correct" ordering \eqref{zordering}, with the resulting evolution equations summing up only the diagrams satisfying such ordering. At the same time, the phase-space integral over the region defined by \eq{Ezordering} is DLA for both the helicity and unpolarized evolutions, as one can easily check explicitly. Hence, for both the unpolarized \cite{Balitsky:1995ub,Balitsky:1998ya,Kovchegov:1999yj,Kovchegov:1999ua,Jalilian-Marian:1997dw,Jalilian-Marian:1997gr,Weigert:2000gi,Iancu:2001ad,Iancu:2000hn,Ferreiro:2001qy} and helicity \cite{Kovchegov:2015pbl,Kovchegov:2016zex,Kovchegov:2017lsr,Kovchegov:2018znm} evolutions to be valid, the ``inverse" ordering \eqref{Ezordering} should not contribute. Below we will clarify how various diagrammatic cancellations prevent the ``inverse" ordering region in \eq{Ezordering} from contributing to both the unpolarized and helicity evolutions, thus confirming that the LLA and DLA evolutions should follow the standard ordering of Eqs.~\eqref{Eordering} and \eqref{zordering}. 

Note that our analysis here corrects an omission at the end of Sec.~II.D.2 in \cite{Kovchegov:2016zex}, where, in the second to last paragraph, the region \eqref{Ezordering} was incorrectly identified as contributing to helicity evolution (and thus included in the equations derived in \cite{Kovchegov:2015pbl}): the equations derived in \cite{Kovchegov:2015pbl} do not contain this region, and we provide a justification for this below.


\subsection{Unpolarized evolution}

For the case of the unpolarized BK/JIMWLK evolution, the cancellation of the region \eqref{Ezordering} is rather straightforward. Note that the conditions \eqref{Ezordering} imply that 
\begin{align}\label{3=1}
x_{31}^2 \ll \frac{z_2}{z_3} \, x_{21}^2 \ll x_{21}^2,
\end{align}
such that the distance $x_{31}$ is very small. In addition, since $z_3 \gg z_2$, the gluon $3$ cannot be emitted off of the gluon $2$. The diagrams contributing in the region \eqref{Ezordering} are shown in \fig{unpol}. For simplicity we are only analyzing the diagrams in the large-$N_c$ limit; we believe that generalization of our argument beyond the large-$N_c$ limit is straightforward. Similarly, we only consider diagrams where the anti-quark at 0 is a spectator: generalization to the case of interacting anti-quark 0 can also be done easily.

Since the dipole $31$ is small, its interaction with the target can be neglected. This way, the numerators of different diagrams in \fig{unpol} are all identical. The difference, and, hence, the relative weights of the diagrams comes from the energy denominators in LCPT notation. Concentrating on the energy denominators only (cf. \cite{Kovchegov:2007vf}), we obtain  
\begin{subequations}\label{A-K}
\begin{align}
& A \propto \frac{1}{E_2^2 \, E_3^2}, \\
& B = C \propto - \frac{1}{E_2^2 \, E_3^2}, \\
& D = E \propto \frac{1}{E_2^2 \, E_3^2}, \\
& F = G = H = I \propto - \frac{1}{2 \, E_2^2 \, E_3^2}, \\ 
& J \propto \frac{3}{4 \, E_2^2 \, E_3^2}, \\
& K \propto - \frac{1}{4 \, E_2^2 \, E_3^2},
\end{align}
\end{subequations}
where $E_2 = {\un k}_2^2/(2 k_2^-)$ and $E_3 = {\un k}_3^2/(2 k_3^-)$ with $k_2$ and $k_3$ the momenta of the two gluons. Note that the kinematics of \eqref{Ezordering} implies that $k_{3 \, \perp} \gg k_{2 \, \perp}$ and $k_3^- \gg k_2^-$. In arriving at Eqs.~\eqref{A-K} we have also added diagrams with the instantaneous terms. For the virtual diagrams, we have only shown the two diagrams (J and K), which, along with their complex conjugate counterparts, give non-zero LLA contributions. 

\begin{figure*}
\begin{center}
\includegraphics[width= 0.9 \textwidth]{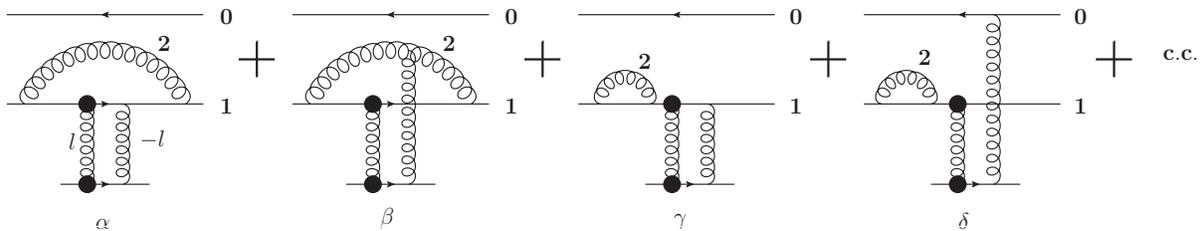} 
\caption{The large-$N_c$ diagrams illustrating emission of unpolarized gluon 2 followed by a Born-level helicity-dependent interaction with the target.}
\label{class_pol}
\end{center}
\end{figure*}

With the help of Eqs.~\eqref{A-K} we see that all the diagrams in \fig{unpol} cancel:
\begin{subequations}\label{unp_canc}
\begin{align}
& A + \thalf \, B + \thalf \, C = 0 , \\
& E + \thalf \, C + G = 0, \\
& D + \thalf \, B + F = 0, \\
& H + I + J + K + \mbox{other virtual diagrams} =0.
\end{align}
\end{subequations}
The grouping of diagram contributions into the individual equations in Eqs.~\eqref{unp_canc} is done to establish resemblance to how the unpolarized small-$x$ evolution emerges from similar cancellations in the standard ordering case of \eqref{Eordering} and \eqref{zordering}. 

We have thus established that, for the unpolarized BK and JIMWLK evolution, the emissions in the kinematics of \eq{Ezordering} do not contribute at the LLA level. This conclusion is in-line with the conventional wisdom. The same cancellation was observed previously in \cite{Iancu:2015vea} (see Fig.~3 there along with the discussion near that figure).


\subsection{Helicity evolution}

Now we consider potential contribution of the ``inverse" light-cone fraction ordering region \eqref{Ezordering} to the case of helicity evolution constructed in \cite{Kovchegov:2015pbl,Kovchegov:2016zex,Kovchegov:2017lsr,Kovchegov:2018znm}. 


\subsubsection{Classical approximation}

Before analyzing the diagrams similar to \fig{unpol} above, let us note that, for the case of helicity, the issue of inverse momentum ordering arises already after one step of evolution due to the dependence of the Born-level initial conditions for the evolution on the center-of-mass energy. This is a simple exercise which would allow us to make a more general conjecture in the operator language, to be verified below. 

Consider a polarized dipole 10, in which we emit one unpolarized gluon 2, such that $z_2 \ll z_1 \approx z_0$. Instead of emitting another gluon 3, let us allow the resulting quark--anti-quark--gluon system to interact with the target, which is taken to be a single polarized quark. The corresponding diagrams are shown in \fig{class_pol}, where, for simplicity, we consider only the case where the quark at 1 is polarized, and the large-$N_c$ limit is applied again.

The difference between the diagrams in \fig{class_pol} is due to different interactions with the target and due to the gluon 2 being emitted and absorbed at different times. The helicity-dependent interaction with the target is indeed energy suppressed. This energy suppression is removed by the energy rescaling factor $z_1 \, s$ (cf. \eq{EQ:Test_operator_pol}). The remaining integral over the transverse momenta of the $t$-channel gluons is logarithmic \cite{Kovchegov:2015pbl,Kovchegov:2016zex}, with the center-of-mass energy squared $s$ providing the UV cutoff. Indeed,  the transverse momentum $l_\perp \equiv |{\un l}|$ of the $t$-channel gluons cannot exceed the center of mass energy of the scattering of the $s$-channel partons on the target. That is, for the diagram $\alpha$ in \fig{class_pol} we have $l_\perp^2 \ll z_1 \, s$, while for the diagram $\beta$ we have $l_\perp^2 \ll \min \{ z_1 \, s, z_2 \, s \} = z_2 \, s$. We have a mismatch, somewhat akin to the opposite ordering problem of \eq{Ezordering}: the interaction of dipole 21 with the target in diagram $\alpha$ ``knows" about the {\sl larger} momentum fraction $z_1$, and does not seem to ``know" of the {\sl smaller} momentum fraction $z_2$ in the dipole 21. At the same time, in the diagram $\beta$ the interaction of the dipole 21 with the target ``knows" about $z_2$.  If this conclusion is correct, then further evolution in the dipole 21 in diagram $\alpha$ may also depend on $z_1$ which is much larger than $z_2$. This is similar to the emission of the (longitudinally) harder gluon 3 after gluon 2 in \fig{gluon23}: this never happens in the standard small-$x$ evolution.   

Luckily we can show that this dangerous contribution cancels. To see this, let us consider the contributions of the graphs in \fig{class_pol}, concentrating only on the $l_\perp$-integral, the overall sign, and the Fourier exponentials for the transform into transverse coordinate space. We write 
\begin{subequations}\label{a-d}
\begin{align}
& \alpha \propto \int\limits^{z_1 s} \frac{d^2 l_\perp}{l_\perp^2} , \\
& \beta \propto - \int\limits^{z_2 s} \frac{d^2 l_\perp}{l_\perp^2} \, e^{i {\un l} \cdot {\un x}_{21}}, \\
& \gamma + \gamma^* \propto - \int\limits^{z_1 s} \frac{d^2 l_\perp}{l_\perp^2} , \\
& \delta + \delta^* \propto \int\limits^{z_1 s} \frac{d^2 l_\perp}{l_\perp^2} \, e^{i {\un l} \cdot {\un x}_{01}},
\end{align}
\end{subequations}
with the upper limit of the integral applying to the $l_\perp^2$ integration only and the asterisk denoting complex conjugation. The Fourier factor makes diagrams $\beta$ and $\delta$ insensitive to the center-of-mass energy, since $1/x_{21}^2 \ll z_2 \, s$ and $1/x_{01}^2 \ll z_1 \, s$. Hence, we can replace $z_1 \to z_2$ in the upper limit of the integral in $\delta + \delta^*$. The remaining $z_1$ dependence in $\alpha$ and $\gamma$ cancels since $\alpha + \gamma + \gamma^* =0$. We thus arrive at the following conclusion:
\begin{align}\label{class_canc}
& \alpha + \beta + \gamma + \gamma^* + \delta + \delta^* \\ & \propto \int\limits^{z_2 s} \frac{d^2 l_\perp}{l_\perp^2} \, \left[ \left( 1 - e^{i {\un l} \cdot {\un x}_{21}} \right) - \left( 1 - e^{i {\un l} \cdot {\un x}_{01}} \right) \right]  \notag \\ & \propto G^{(0)} (x_{21}^2, z_2) - G^{(0)} (x_{01}^2, z_2). \notag
\end{align}
(In the last line we have introduced the Born-level polarized dipole amplitude \cite{Kovchegov:2015pbl,Kovchegov:2016zex}
\begin{align}\label{G0_init}
G^{(0)} (x_{01}^2, z_2)&  = - \as^2 \, \frac{C_F}{N_c} \, \int\limits^{z_2 s} \frac{d^2 l_\perp}{l_\perp^2} \, \left( 1 - e^{i {\un l} \cdot {\un x}_{01}} \right) \notag \\ &  = - \as^2 \, \frac{C_F}{N_c} \, \pi \, \ln (z_2 \, s \, x_{01}^2),
\end{align}
representing the contribution of the helicity-dependent part of the two $t$-channel gluons exchange.) 

\begin{figure}
\begin{center}
\includegraphics[width= 0.45 \textwidth]{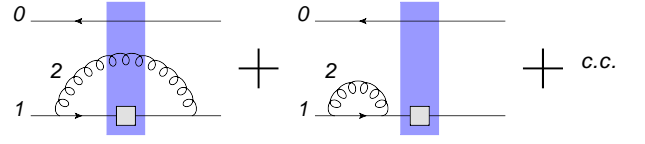} 
\caption{A step of the polarized dipole evolution due to emission of a soft unpolarized gluon.}
\label{conj}
\end{center}
\end{figure}

We observe that $z_1$ dependence disappeared from \eqref{class_canc}. Therefore, the interaction of the dipole 21 with the target does not depend on $z_1$, and can be thought as governed by the smaller momentum fraction $z_2$.\footnote{Note that the dependence on $z_2$ also cancels out in \eq{class_canc}. This does not affect our conclusion that further evolution can be though of a depending on $z_2$ only, even if this $z_2$ dependence vanishes at the low order of diagrams considered here in \fig{class_pol}.} While this conclusion is derived here in the ``quasi-classical" case, that is, for Born-level $t$-channel gluon exchanges instead of further small-$x$ evolution (following the emission of gluon 2), our experience and conventional wisdom in small-$x$ physics suggest that such quasi-classical conclusions about operators usually continue to apply when small-$x$ evolution corrections are included. Therefore, we make the following conjecture, which we believe to be valid beyond the specific case of \fig{class_pol}, applying to the general helicity evolution case pictured in \fig{conj}: after emitting the soft gluon 2, the Wilson line of quark 1, when interacting with the shock wave, should be replaced by the same Wilson line, but dependent on $z_2$,
\begin{align}
W_{\un 1}^{(R)} (z_1 - z_2) \to W_{\un 1}^{(R)} (z_2).
\end{align}
The dependence of Wilson lines on the light-cone momentum fractions comes in through the cutoffs on the appropriate integrals, which can be implemented in a variety of different ways \cite{Balitsky:2008zza}. We have used $W^{(R)}$ to denote the Wilson line at ${\un x}_1$, to stress that our conjecture should be valid for a line in any irrep $R$. 

\subsubsection{Two steps of evolution: the $s$-channel diagrams}

\begin{figure*}
\begin{center}
\includegraphics[width= 0.9 \textwidth]{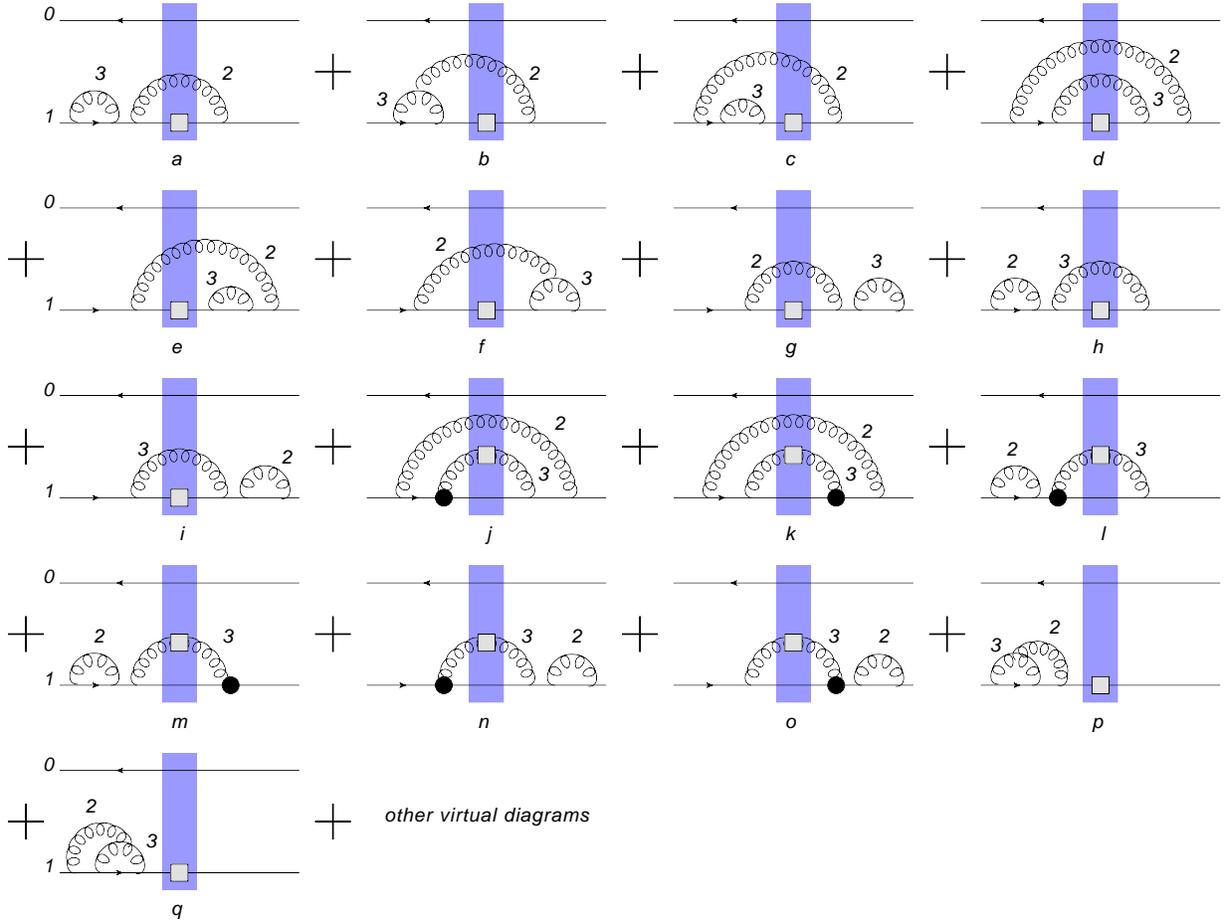} 
\caption{The large-$N_c$ diagrams contributing to emission of two gluons (2,3) in a polarized dipole 10 in the kinematics of \eq{Ezordering} for helicity evolution.}
\label{polarized}
\end{center}
\end{figure*}

Now, let us test the above-stated conjecture by considering small-$x$ evolution corrections. The DLA diagrams contributing to emission of gluons 2 and 3 in the polarized dipole 10, in the kinematics of \eq{Ezordering} are shown in \fig{polarized}. We again assume the large-$N_c$ limit for simplicity. Once more, not all the virtual diagrams are shown explicitly: we show only the virtual diagram which are DLA (along with their counterparts obtained by complex conjugation). While the gluon 2 is again assumed to be unpolarized, the gluon 3 may or may not be polarized. To simplify our analysis, we did not include the splittings leading to creation of a soft quark and a hard gluon \cite{Kovchegov:2015pbl}: these can be considered separately (or the quark lines in \fig{polarized} can be thought of as originating in the gluon lines of an adjoint polarized dipole 10 in the large-$N_c$ limit). 

Throughout this Subsection, we will be interested in helicity-dependent contributions only: when evaluating diagrams, it will be understood that only such contributions are included. 

Employing a similar analysis to what was done for the diagrams in \fig{unpol}, we observe the following cancellations among the diagrams in \fig{polarized}:
\begin{subequations}
\begin{align}
& a + b  +\thalf \, c = 0 , \\
& \thalf \, e + f + g =0, \\
& h + i + d = 0.
\end{align}
\end{subequations}
We are left with the diagrams $j$ through $q$, along with $\thalf \, c$ and $\thalf \, e$. To find their contribution, we remember that emission of a polarized gluon comes with an addition factor of $\thalf$ compared to the emission of an eikonal gluon (compare Eqs.~\eqref{Keik2} and \eqref{Kbeta}). In addition, the polarized Wilson line in the adjoint representation comes with the factor $\eta_A = 2$, while the polarized Wilson line in the fundamental representation has $\eta_F = 1$ (see \eq{eta_def}). With this in mind, one can show that 
\begin{align}\label{pol_sum1}
& j + k + \ldots + q + p^* + q^* + \thalf (c+e)  \\ & \propto 2 \, G_{31} + 2 \, G_{32} -  2 \, G_{31} - 2 \, G_{30} - G_{12} + G_{10} \notag \\ & \approx G_{12} - G_{10}, \notag 
\end{align}
where, in the last step, we have used the fact that ${\un x}_3 \approx {\un x}_1$ in the kinematics of \eq{Ezordering} (see \eq{3=1}).

In \eq{pol_sum1} we have used the definition of the polarized dipole amplitude \cite{Kovchegov:2018znm}
\begin{align}\label{Gdef10}
G_{10} (z) = \frac{z_{pol} \, s}{N_c}  \:  \mbox{Re} \:  \left\langle \mbox{T} \, \mbox{tr} \left[ V_{\ul 0} \, V_{{\un 1}}^{pol \, \dagger} \right] + \mbox{T} \, \mbox{tr} \left[ V_{{\un 1}}^{pol} \, V_{\ul 0}^\dagger \right] \right\rangle ,
\end{align}
where $z_{pol}$ is the light-cone momentum fraction of the polarized line, while $z$ is the smaller one of the momentum fractions of the two lines. The object in \eq{G0_init} is the impact-parameter integrated polarized dipole amplitude,
\begin{align}
G(x_{10}^2, z) \equiv \int d^2 x_1 \, G_{10} (z) .
\end{align}

\begin{figure*}
\begin{center}
\includegraphics[width= 0.75 \textwidth]{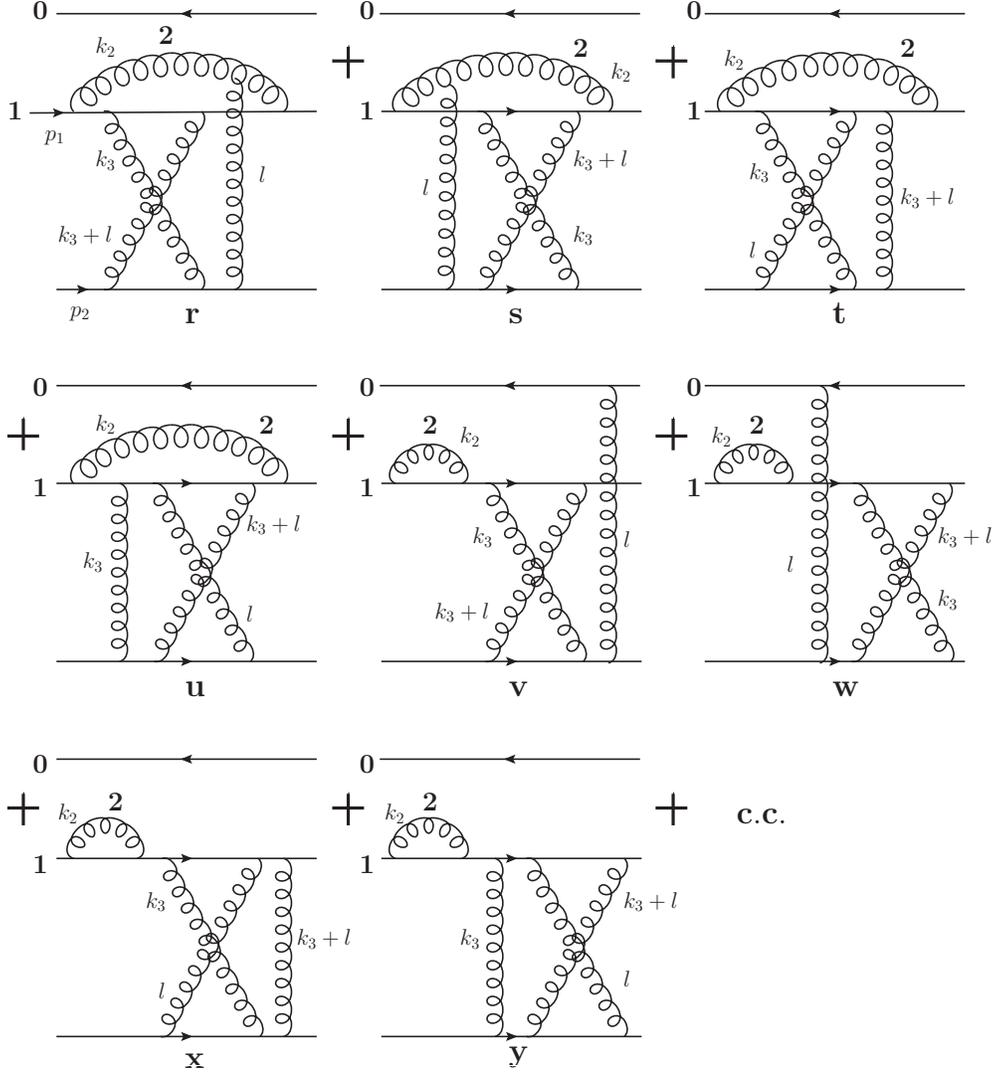} 
\caption{Shock wave correction diagrams in the large-$N_c$ limit. The complex conjugate (c.c.) operation applies to the diagrams with the virtual gluon 2 only.}
\label{shock_wave}
\end{center}
\end{figure*}

We see that, unlike the unpolarized case considered in the previous Section, the $s$-channel shock wave diagrams do not cancel in the inverse ordering regime of \eq{Ezordering}. In retrospect, this may seem natural. In the unpolarized evolution, the result \eqref{unp_canc} was due to real-virtual cancellation combined with the observation that if a quark emits a collinear gluon, forming a very close quark-gluon pair, the whole pair still interacts with the shock wave as a single quark. This is because, in the unpolarized eikonal case, the shock wave only couples to the net color of the compact object. In the helicity evolution case, the shock wave, at the sub-eikonal level, couples to the helicity of objects scattering on it. Helicity is not conserved (unlike the net angular momentum), which does not allow for the same real-virtual cancellations as in \eq{unp_canc}. In addition, virtual polarized gluon correction is not DLA  \cite{Kovchegov:2015pbl}, which also prevents the real-virtual cancellations from taking place. In the unpolarized evolution, the real-virtual cancellations ensure that the Balitsky--Fadin--Kuraev--Lipatov (BFKL)
\cite{Kuraev:1977fs,Balitsky:1978ic}, BK and JIMWLK evolution equations are UV finite. In the case of helicity evolution \cite{Kovchegov:2015pbl}, the real-virtual non-cancellation makes the transverse position integrals logarithmically divergent in the UV, and this divergence is regularized by the (inverse) center-of-mass energy, thus creating the DLA resummation parameter $\as \, \ln^2 (1/x)$ which is absent in the unpolarized evolution. Hence, the non-cancellation we see in \eq{pol_sum1} appears to be intimately tied to the generation of DLA resummation parameter employed in helicity evolution. Still, for the calculation at hand this non-cancellation presents a problem. Since the non-cancellation \eqref{pol_sum1} appears to take place due to the more ``sophisticated" non-eikonal interactions with the shock wave, it seems reasonable to try to see whether shock wave corrections may arise to cancel the contribution of the undesirable region \eqref{Ezordering}.

To verify this hypothesis, let us consider a specific contribution where the interaction with the shock wave in \fig{polarized} is due to the Born-level two-gluon exchange only. Reinstating the $(x \cdot y)$ factors from Eqs.~\eqref{Keik2} and \eqref{Kbeta} we re-write \eq{pol_sum1} as an equality (in the large-$N_c$ limit),
\begin{align}\label{pol_sum2}
& j + k + \ldots + q + p^* + q^* + \thalf (c+e)  \\ & = \left( \frac{\as \, N_c}{2 \pi} \right)^2 \int\limits_{\Lambda^2/s}^1 \frac{d z_2}{z_2} \,  \int\limits_{z_2}^1 \frac{d z_3}{z_3} \, \int\limits_{1/z_2 s}^{x_{10}^2} \frac{d x_{21}^2}{x_{21}^2} \int\limits_{1/z_3 s}^{(z_2/z_3) \,  x_{21}^2} \frac{d x_{31}^2}{x_{31}^2} \notag \\ & \times \left[ G^{(0)} (x^2_{12} , z_2) - G^{(0)} (x^2_{10} , z_2) \right]. \notag
\end{align}
In arriving at \eq{pol_sum2} we have also integrated over impact parameters (e.g., over ${\un x}_1$). The explicit expression for $G^{(0)}$ is given in \eq{G0_init}.

The IR cutoff of $x_{10}^2$ on the $x_{21}^2$ integral results from summing diagrams with the gluon 2 emitted or absorbed by the anti-quark at 0 as well. This cutoff is not essential for the calculations below.


\subsubsection{Two steps of evolution: shock wave corrections}

We will work in the $A^- =0$ light-cone gauge of the projectile. In this gauge, the shock wave correction diagrams potentially giving DLA contributions can be obtained from the graphs in \fig{class_pol} by including another gluon connecting the projectile and the target. The shock wave corrections giving DLA contributions in the large-$N_c$ limit are shown in \fig{shock_wave}. We do not show the non-DLA shock wave corrections diagrams explicitly, since there are more of these diagrams than the DLA shock-wave correction ones. 

Once again the quark line at 1 is polarized. However, we do not show explicitly which quark-gluon vertices are sub-eikonal (and, hence, transfer polarization), since in each diagram it could be either the two vertices attached to the $k_3$ or to the $k_3 +l$ gluon lines --- both contributions are included in the calculation. Since the target is taken to be a single quark, and the polarized line at 1 is a quark, the large-$N_c$ diagrams of \fig{shock_wave} do not appear to be planar. 

The quark at 1 and the quark in the target enter the scattering process with large momenta $p_1^-$ and $p_2^+$ (see the diagram $r$ in \fig{shock_wave} for these momenta labels). We perform the calculation in the following approximation:
\begin{align}\label{assum1}
p_1^-, p_2^+ \gg k_3^+, k_3^-, k_{3 \, \perp}, k_2^+, k_2^-, k_{2 \, \perp}, l^+, l^-, l_\perp \gg p_1^+, p_2^- .
\end{align}
In addition, we impose the kinematics of \eq{Ezordering},
\begin{align}\label{assum2}
k_3^- \gg k_2^-, \ \ \ \frac{{\un k}_3^2}{2 k_3^-} \gg \frac{{\un k}_2^2}{2 k_2^-} .
\end{align}
Among other things, this prevents the $k_3$ and $k_3 +l$ gluon lines from connecting to the $k_2$ gluon line, thus limiting the number of diagrams.

The momentum labeling in the diagrams of \fig{shock_wave} is not random, and demonstrates the diagrammatic phase space which generates the DLA contributions. For instance, while all kinematic assumptions are shown above in Eqs.~\eqref{assum1} and \eqref{assum2}, the triple gluon vertex in, say, diagrams $r$ and $s$ is eikonal only for $k_2^- \gg |l^-|$, resulting in the $k_3^- \gg k_2^- \gg |l^-|$ ordering of light-cone momenta. In other diagrams of \fig{shock_wave}, the $k_2^- \gg |l^-|$ ordering results from the poles in the calculation after the $k_3^- \gg |l^-|$ assumption was imposed. 

Employing the same proportionality factor as in \eq{pol_sum1}, after a rather tedious calculation one arrives at the following results in the DLA and in the large-$N_c$ limit:
\begin{subequations}\label{pol_sum3}
\begin{align}
r = s \propto - \thalf \left[ G^{(0)} (x^2_{32} , z_2) - G^{(0)} (1/\Lambda^2 , z_2) \right], \\
t = u \propto  \thalf \left[ G^{(0)} (x^2_{31} , z_2) - G^{(0)} (1/\Lambda^2 , z_2) \right], \\
v = w \propto \tfrac{1}{4} \left[ G^{(0)} (x^2_{30} , z_2) - G^{(0)} (1/\Lambda^2 , z_2) \right], \\
x = y \propto - \tfrac{1}{4} \left[ G^{(0)} (x^2_{31} , z_2) - G^{(0)} (1/\Lambda^2 , z_2) \right],
\end{align}
\end{subequations}
with $\Lambda$ an IR cutoff, as before. Once again we stress that only the helicity-dependent part of the diagrams is considered here, that is, the part proportions to $\sigma_1 \, \Sigma$ with $\sigma_1$ the helicity of quark 1 and $\Sigma$ the helicity of the target. Combining all the shock-wave correction diagrams we arrive at
\begin{align}\label{pol_sum4}
& r+s+ \ldots +y +v^* +w^* +x^* +y^* \notag \\ & \propto - G^{(0)} (x^2_{32} , z_2) + G^{(0)} (x^2_{30} , z_2) \notag \\ & \approx - G^{(0)} (x^2_{12} , z_2) + G^{(0)} (x^2_{10} , z_2).
\end{align}
Inserting all the missing integrations and prefactors, we rewrite \eq{pol_sum4} as an equality,
\begin{align}\label{pol_sum5}
&  r+s+ \ldots +y +v^* +w^* +x^* +y^* \notag \\ & = - \left( \frac{\as \, N_c}{2 \pi} \right)^2 \int\limits_{\Lambda^2/s}^1 \frac{d z_2}{z_2} \,  \int\limits_{z_2}^1 \frac{d z_3}{z_3} \, \int\limits_{1/z_2 s}^{x_{10}^2} \frac{d x_{21}^2}{x_{21}^2} \int\limits_{1/z_3 s}^{(z_2/z_3) \,  x_{21}^2} \frac{d x_{31}^2}{x_{31}^2} \notag \\ & \times \left[ G^{(0)} (x^2_{12} , z_2) - G^{(0)} (x^2_{10} , z_2) \right]. 
\end{align}
We see that, indeed, the shock wave corrections cancel the inverse ordering contribution \eqref{pol_sum2},
\begin{align}\label{pol_canc5}
& j + k + \ldots + q + p^* + q^* + \thalf (c+e) \\ & + r+s + \ldots +y +v^* +w^* +x^* +y^* = 0. \notag
\end{align}
We conclude that the diagrams in the kinematics of \eq{Ezordering} cancel. Hence, the inverse ordering diagrams do not contribute to our helicity evolution. 

\begin{figure*}
\begin{center}
\includegraphics[width= 0.9 \textwidth]{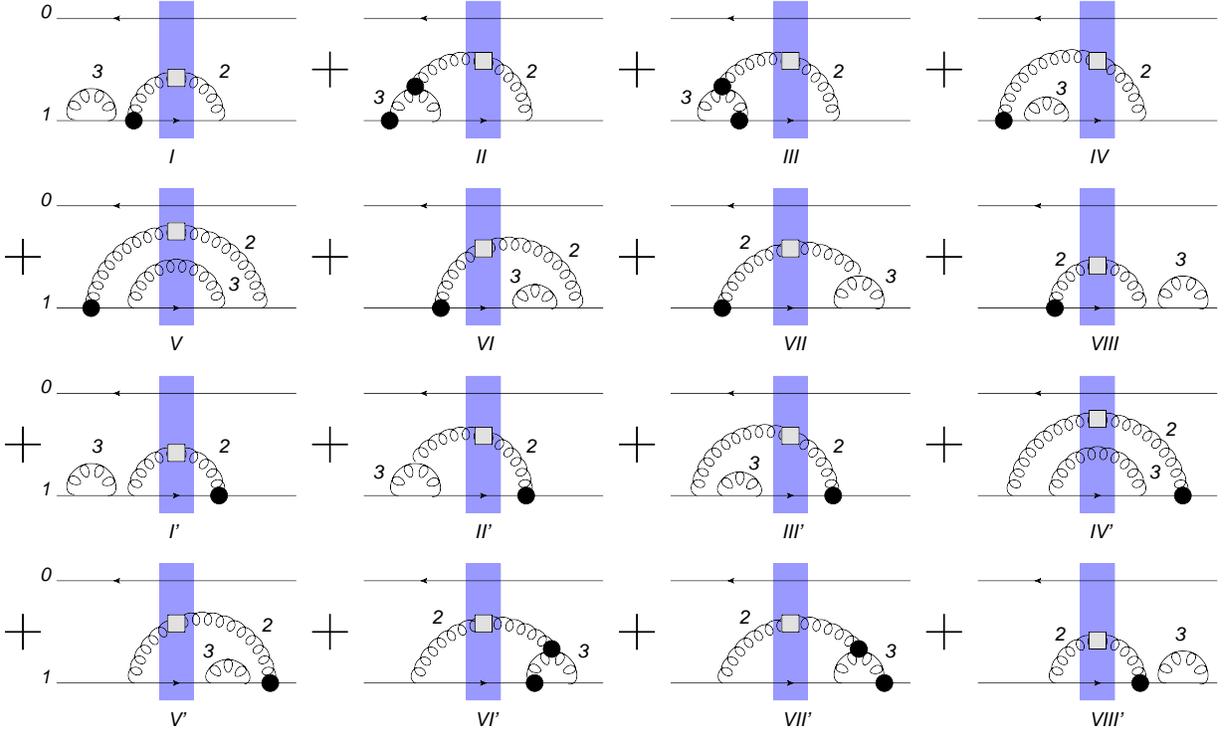} 
\caption{The large-$N_c$ diagrams contributing to emission of two gluons (2,3) in a polarized dipole 10 in the kinematics of \eq{Ezordering} for helicity evolution, now with the gluon 2 polarized.}
\label{polarized2}
\end{center}
\end{figure*}


\subsubsection{Early emission of the polarized gluon}

For completeness, we should also consider the case when the gluon 2 is polarized in the inverse-ordering kinematics \eqref{Ezordering}. Our discussion here will be brief, following the steps outlined above. The $s$-channel diagrams contributing to the emission of gluons 2 and 3 in the polarized dipole 10 in the large-$N_c$ limit with the kinematics of \eqref{Ezordering} and with the gluon 2 being polarized are shown in \fig{polarized2}. 

By analogy to Eqs.~\eqref{unp_canc}, we observe the following cancellations:
\begin{subequations}
\begin{align}\label{pol_canc7}
\thalf \, VI + VII + VIII = 0, \\
I' + II' + \thalf \, III' =0, \\
\thalf \,  IV + V + \thalf \, VI =0, \\
\thalf \, III' + IV' + \thalf \, V' =0, \\
I + III + \thalf \, IV =0, \\
\thalf \, V' + VI' + VIII' = 0. 
\end{align}
\end{subequations}
We are left with the diagrams $II$ and $VII'$. Their contribution is   
\begin{align}\label{pol_sum8}
&  II + VII' \notag \\ & = \left( \frac{\as \, N_c}{2 \pi} \right)^2 \int\limits_{\Lambda^2/s}^1 \frac{d z_2}{z_2} \,  \int\limits_{z_2}^1 \frac{d z_3}{z_3} \, \int\limits_{1/z_2 s}^{x_{10}^2} \frac{d x_{21}^2}{x_{21}^2} \int\limits_{1/z_3 s}^{(z_2/z_3) \,  x_{21}^2} \frac{d x_{31}^2}{x_{31}^2} \notag \\ & \times 2 \, \left[ G^{(0)} (x^2_{12} , z_2) + G^{(0)} (x^2_{20} , z_2) \right]. 
\end{align}
Again we assume that the interaction with the shock wave in \fig{polarized2} is a Born-level exchange of two $t$-channel gluons. 

\begin{figure}
\begin{center}
\includegraphics[width= 0.45 \textwidth]{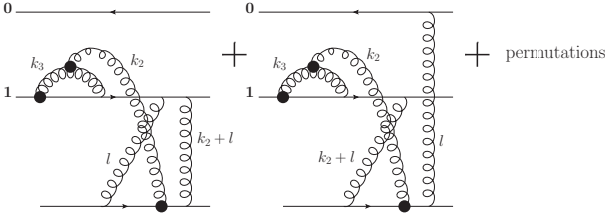} 
\caption{The types of shock wave diagrams that should cancel the contributions of the diagrams II and VII' in \eq{pol_sum8}.}
\label{shock_wave3}
\end{center}
\end{figure}

Based on the above calculations, we expect the remaining diagrams II and VII' from \fig{polarized2} to be canceled by the diagrams of the type shown in \fig{shock_wave3} in the kinematics of \eqref{Ezordering}. 

\begin{figure}[h]
\begin{center}
\includegraphics[width= 0.45 \textwidth]{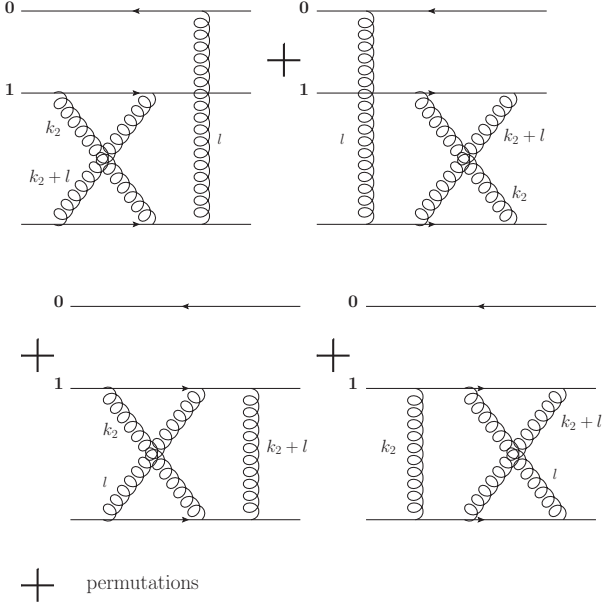} 
\caption{Shock wave corrections to the dipole amplitude at one step of DLA evolution.}
\label{shock_wave4}
\end{center}
\end{figure}


\subsubsection{Remaining shock-wave corrections}

The above conclusion that shock wave corrections could be DLA may still be troubling: one may worry whether similar shock wave corrections could affect the evolution in the standard ordering \eqref{zordering}. This is a legitimate concern: above, the inverse ordering condition $k_3^- \gg k_2^-$ was important for the selection of the shock-wave correction diagrams that contribute (e.g., line $k_3$ does not connect to $k_2$ in \fig{shock_wave}), but was not important in the actual diagram evaluation. If we instead consider the standard momentum ordering condition $k_2^- \gg k_3^-$, the shock wave corrections in \fig{shock_wave} appear to still contribute, while the diagrams in \fig{shock_wave3} become impossible and can be discarded. The issue of persistence of the shock-wave correction diagrams in \fig{shock_wave} can be already seen at one step of the polarized dipole evolution, as illustrated in \fig{shock_wave4}. The diagrams in \fig{shock_wave4} can be obtained by considering the diagrams in \fig{shock_wave} and removing the gluon 2 in the latter. Permutations in \fig{shock_wave4} label the diagrams where either the gluon $k_2$ or the gluon $k_2 +l$ connect to line 0. The diagrams in  \fig{shock_wave4}, along with the permutations, combine to give (in the large-$N_c$ limit)
\begin{align}\label{init_sum1}
\frac{\as \, N_c}{2 \pi} \!\!\!\!\! \int\limits_{1/(s x_{10}^2)}^z \!\!\!  \frac{d z_2}{z_2} \,   \int\limits_{1/z_2 s}^{x_{10}^2} \frac{d x_{21}^2}{x_{21}^2}  \left[ G^{(0)} (x^2_{12} , z_2) - G^{(0)} (x^2_{20} , z_2) \right]. 
\end{align}
It appears that this contribution is not canceled by any other shock-wave correction diagram (or by any other diagram not included in one step of evolution). Hence, it has to be included in the final answer. Below we present a way this contribution can be incorporated in helicity evolution based on the existing calculations: our results below are tentative and will have to be verified by a detailed calculation of the shock wave corrections with the standard momentum ordering  \eqref{zordering} in the future. 

The diagrams in \fig{shock_wave4} can be redrawn as being due to emissions of the long-lived (in $x^-$ direction) gluons $k_2$ and $k_2 + l$, whose emission can be described by the kernel $\calk^\beta$. Since factorization of the diagrams in \fig{shock_wave4} into the long-lived gluons $k_2$ and $k_2 + l$ and a quick interaction with the target is possible, it appears that the contribution of the diagrams in \fig{shock_wave4} can be included into helicity evolution by modifying the initial conditions. Compare the contribution \eqref{init_sum1} to one step of the large-$N_c$ evolution for the polarized dipole amplitude, where the evolution is generated only by the emission of the polarized gluon (that is, by the application of $\calk^\beta$ kernel from \eqref{Kbeta}, see also the first line of Eq.~(80) in  \cite{Kovchegov:2015pbl})
\begin{align}\label{init_sum3}
\frac{\as \, N_c}{2 \pi}  \int\limits_{1/(s x_{10}^2)}^z \frac{d z_2}{z_2} \,   \int\limits_{1/z_2 s}^{x_{10}^2} \frac{d x_{21}^2}{x_{21}^2}  \left[ 2 \, G_{12} (z_2) + 2 \, \Gamma_{20,21} (z_2) \right]. 
\end{align}
Here we have employed the ``neighbor" polarized dipole amplitude $\Gamma_{20,21} (z_2)$, whose definition can be found in \cite{Kovchegov:2015pbl,Kovchegov:2016zex,Kovchegov:2017lsr,Kovchegov:2018znm}. We see  that the contribution \eqref{init_sum1} can be generated by the evolution \eqref{init_sum3} if $G_{12} (z_2) = \tfrac{1}{2} G^{(0)}_{12} (z_2)$ and $\Gamma_{20,21} (z_2) = - \thalf G^{(0)}_{20} (z_2)$. (Note that \eq{init_sum1} is integrated over impact parameters, while \eq{init_sum3} is written for the fixed impact parameter.) It appears that we need to add the following corrections to the inhomogeneous terms of the equations for $G$ and $\Gamma$:
\begin{align}\label{init_sum4}
\delta G^{(0)}_{12} (z_2) = \tfrac{1}{2} \, G^{(0)}_{12} (z_2), \ \ \ \delta \Gamma_{20,21} (z_2) = - \thalf \, G^{(0)}_{20} (z_2). 
\end{align}
For completeness, let us quote the resulting large-$N_c$ helicity evolution equations:
\begin{widetext}
\begin{subequations}\label{init_sum5}
\begin{align}
& G_{10} (z) = \frac{3}{2} \, G_{10}^{(0)} (z) + Q_{10}^{(0)} (z) + \frac{\alpha_s \, N_c}{2 \pi} \int\limits_{\frac{1}{s \, x_{10}^2}}^{z}
\frac{dz'}{z'} \int\limits^{x_{10}^2}_\frac{1}{z' s} \frac{d x_{21}^2}{x_{21}^2} \: \left[ \Gamma_{10,21} (z') + 3 \, G_{21} (z')  \right], \\
& \Gamma_{10,21} (z') = \frac{1}{2} \, G_{10}^{(0)} (z) + Q_{10}^{(0)} (z') + \frac{\alpha_s \, N_c}{2 \pi} \int\limits_{\min \{ \Lambda^2, \frac{1}{x_{10}^2} \} / s }^{z'}
\frac{dz''}{z''} \int\limits^{\min \{ x_{10}^2, x_{21}^2 z'/z'' \} }_\frac{1}{z'' s} \frac{d x_{32}^2}{x_{32}^2} \: \left[ \Gamma_{10,32} (z'') + 3 \, G_{32} (z'')  \right].
\end{align}
\end{subequations}
\end{widetext}
Here, $Q_{10}^{(0)} (z)$ is the contribution to the inhomogeneous term due to Born-level exchange of $t$-channel quarks. (A more detailed calculation may be needed to determine whether this quark exchange contribution also gets modified by some diagrams analogous to those in \fig{shock_wave4}.) Note that the small-$x$ asymptotics is unaffected by the exact form of the inhomogeneous terms in Eqs.~\eqref{init_sum5} \cite{Kovchegov:2016weo}: hence, the modifications of the inhomogeneous terms in Eqs.~\eqref{init_sum5} would not change the small-$x$ asymptotics of helicity distributions derived in \cite{Kovchegov:2016weo,Kovchegov:2017jxc,Kovchegov:2017lsr}. The inhomogeneous terms in the large-$N_c \& N_f$ equations \cite{Kovchegov:2015pbl,Kovchegov:2018znm} have to be modified accordingly as well. The diagrams in \fig{shock_wave4} should also be taken into consideration when constructing the inhomogeneous term $\mathcal{W}_\tau^{(0) \, pol}$ for the helicity JIMWLK derived in the main text (to be further detailed in \cite{CK2}). Since we did not perform a full calculation of the shock wave corrections for the standard momentum ordering ($k_2^- \gg k_3^-$), our conclusions in this Subsection on how the inhomogeneous terms get modified are only tentative.

Finally, let us make the following observation: in the $A^- =0$ gauge, the interaction with the target always struggles to be eikonal. Eikonal quark-gluon vertices in the ``+"-moving target come in with $\gamma^+$, which couples to the $A^-$ component of the gluon field, which is zero in the $A^- =0$ gauge. The leading gluon field component in this gauge is the transverse field, ${\un A} (x)$. Unpolarized interaction of two $A_\perp$ gluons with the target is eikonal due to pinching of the $1/l^-$ pole in the gluon propagators. Note that after this pole pinching the $l^-$ integration is already carried out and cannot give us a logarithm of energy.  Interaction of three $A_\perp$ gluons with the target, as pictured in \fig{shock_wave4}, includes the pinching of the $1/l^-$ pole, along with the un-pinched integral over $k_2^-$, which gives a logarithm of energy.  Due to the absence of pinching in the   $k_2^-$ integral, the contribution is sub-eikonal, as expected for helicity. We see that the direct interaction of the gluons with the target can be of two types: it can either be eikonal due to pinching (but not logarithmic), or it can be energy-suppressed, and bring in a logarithm of energy. Therefore, if we are doing our calculation at the sub-eikonal level, we can only have one $k_2^-$-type gluon interacting with the target, bringing in energy suppression and a logarithm of energy: more than three exchanges of such logarithm-generating gluons with the target are further suppressed in energy, and are sub-sub-eikonal and beyond. Additional exchanges of pinched gluons, beyond those at the leading order, do not bring logarithms of energy, and are not DLA. Hence, at the sub-eikonal level of this helicity calculation, and at DLA, we cannot have more than three  $A_\perp$ gluons interacting with the target directly. This means that the diagrams in \fig{shock_wave4} are the only types of direct interaction of more than two $A_\perp$ gluons with the target possible in our sub-eikonal helicity calculation. By including them in the modification of the initial conditions for our evolution we accounted for all such target-interaction contributions.

~\\~~\\

We conclude this Appendix by summarizing one more time that we have observed cancellations between the $s$-channel diagrams and shock-wave corrections in the inverse ordering kinematics \eqref{Ezordering} for helicity evolution. In the standard small-$x$ evolution ordering of lifetimes \eqref{Eordering} and light-cone momentum fractions \eqref{zordering} the shock wave corrections appear to contribute, and can be absorbed into the inhomogeneous terms for the evolution. These results present one of the most stringent cross-checks to date of the helicity evolution equations derived in \cite{Kovchegov:2015pbl,Kovchegov:2016zex,Kovchegov:2017lsr,Kovchegov:2018znm} and employed in the main text here.



%

\end{document}